\documentclass[tikz,12pt]{article}
\usepackage{mheckII}

\usepackage[T1]{fontenc}

\usepackage{tikz-feynman}

\usepackage{manfnt}
\usepackage{longtable}

\usepackage{fancybox}

\usepackage{multirow}

\usepackage{blkarray}

\usepackage{tikz} 
\usetikzlibrary{matrix}

\theoremstyle{theorem}

\newtheorem{fact}{Fact}

\newtheorem{claim}{Claim}

\newtheorem*{phm}{Physical Mechanism}
\newtheorem*{ped}{Pedantic Version}
\newtheorem*{bid}{Suggestion}

\theoremstyle{definition}
\newtheorem{defn}{Definition}

\newtheorem{rem}{Remark}

\newtheorem*{exe}{Example}
\newtheorem*{exe/c}{Example/Claim}

\def\Dsl{\,\raise.15ex\hbox{/}\mkern-13.5mu D}
\def\dsl{\,\raise.25ex\hbox{/}\mkern-10.5mu \partial}

\renewcommand{\arraystretch}{1.5}

\title{Hwang-Oguiso invariants and frozen singularities
in Special Geometry
}

\authors{Sergio Cecotti\footnote{e-mail: {\tt cecotti@sissa.it}, {\tt cecotti@bimsa.cn}}\vskip 9pt

\centerline{Yanqi Lake Beijing Institute of Mathematical Sciences and Applications (BIMSA)
}
\centerline{Yanqi Island, Huairou District, Beijing 101408, China}
}

\abstract{In Special Geometry there are two inequivalent notions of ``Kodaira type'' for a singular fiber: one associated with its local monodromy and one with its Hwang-Oguiso characteristic cycle. When the two Kodaira types are \emph{not equal} the geometry is subtler and its deformation space gets smaller (``partially frozen'' singularities). The paper analyzes the physical interpretation of the Hwang-Oguiso invariant in the context of 4d $\cn=2$ QFT
and describes the surprising phenomena which appear when it does not coincide with the monodromy type.
The Hwang-Oguiso multiple fibers are in one-to-one correspondence with the partially frozen singularities in M-theory compactified on a local elliptic K3: a chain of string dualities relates the two geometric set-ups. Paying attention to a few subtleties, this correspondence explains in purely geometric terms how the ``same'' Kodaira elliptic fiber may have different deformations spaces. The geometric computation of the number of deformations agrees with the physical expectations. At the end we briefly outline the implications of the Hwang-Oguiso invariants for the classification program of 4d $\cn=2$ SCFTs.}

\begin{document}
\maketitle

\tableofcontents


\section{Introduction: freezing singularities and all that}

There is a number of supersymmetric complex geometries (such as compactifications of F-theory \cite{F1,F2,F3}, Seiberg-Witten geometries \cite{SW1,SW2,M1,M2}, etc.)
where the possible local monodromies around codimension-1 singularities are in 1-to-1 correspondence with Kodaira's singular elliptic fibers \cite{koda1,koda2,koda3,bhm}. One says that the codimension-1
 singularity has the ``Kodaira type'' of its local monodromy.
 The local monodromy is expected to control the physics in the vicinity of the singularity, but there is the ``surprising'' phenomenon that some Kodaira types come in several versions
with distinct physics and deformation spaces of different dimensions.  
The version with most deformations is called ``unfrozen'' and the other ones ``(partially) frozen'' \cite{F1,F2,F3,MM1,MM2,YYY}.
The physical distinction between the various versions is well understood, 
but the geometrical side still looks a bit mysterious. Some authors simply think that 
analytic geometry \emph{per se} is too a coarse tool to capture the full complexity of SUSY quantum systems: hence the same geometry may be associated to several systems
with different physics. While this sounds reasonable, it cannot explain the variability of the deformation space since one should be able to study the deformations of the geometry by
purely geometric means.   

In this paper we discuss \emph{geometrically} the distinct singularities with the 
same local monodromy in the context of
(rigid) \emph{special geometry} i.e.\! the integrable systems \cite{D1,D2} associated to the Seiberg-Witten geometry of 4d $\cn=2$ SCFTs and some natural generalizations of them
(see \S.\,\ref{s:preliminary} for definitions).
Although this may look a simplified context,  it turns out that these geometries are related by string dualities
to frozen singularities in M- and F-theory.

For clarity we distinguish two ``freezing mechanisms'' with somehow different physical interpretations.
At a deeper level they share the same mathematical origin and follow from the same physical idea. We stress that the two mechanisms may operate simultaneously: in this case the resulting singular geometry is pretty subtle
(yet fully understood).

In the rest of this introduction we give an informal survey of the mathematics and physics beyond the freezing and sketch the relation with M- and F-theory.

\subparagraph{The geometric mechanism.}
Special geometry is (in particular) a holomorphic fibration over a base of dimension $r$
whose generic fibers are Abelian varieties of complex dimension $r$.
When $r=1$ this is an elliptic fibration over a curve;
the singular fibers which may arise\footnote{\ Multiple fibers cannot arise in an elliptic fibration which is Lagrangian.} in codimension-$1$ were classified by Kodaira
\cite{koda1,koda2,koda3}\footnote{\ We review the ellptic story in \S.\,\ref{s:rank1} for convenience of the reader.}
see tables \ref{table} and \ref{table2}. To understand special geometry in rank $r\geq2$ one needs to extend Kodaira's analysis to higher dimension.
In doing this one may follow different strategies. The simplest one is to use the local monodromy
$\varrho_\alpha\in SL(2,\Z)\subset Sp(2r,\Z)$ as the guiding principle:
one declares the singular fiber to be the higher dimensional analogue of the Kodaira fiber with the same conjugacy class in $SL(2,\Z)$. This is the physicists's way to assign ``Kodaira types'' to higher dimensional singular fibers in codimension-$1$.

A more sophisticate strategy is to look for a maximal connected cycle $\boldsymbol{\Psi}\equiv \sum_i m_i \Theta_i$ of rational curves $\Theta_i$ inside the singular fiber. The intersection matrix $\Theta_i\cdot\Theta_j$ is the negative of
the Cartan matrix of an $\tilde A\tilde D\tilde E$ affine Lie algebra\footnote{\ The zero and the $A^\infty_\infty$ matrices are also allowed.} while the reduced multiplicities $m_i/\!\gcd(m_j)$ are the Coxeter labels of the nodes in the affine Dynkin graph. Hwang and Oguiso say that the higher dimensional fiber is \emph{modelled} on the Kodaira fiber with the
same (reduced) cycle $\boldsymbol{\Psi}/\!\gcd(m_i)$, and assign to it the Kodaira type of the  reduced cycle \cite{oguiso1,oguiso2,oguiso3}.

We end up with two different notions of ``Kodaira type'' for the singular fiber: the monodromy one and the Hwang-Oguiso one. When the two agree, everything works as naively expected:
 we shall show in \S.\,\ref{s:simplefib} that in this case the singular fiber is just a product of the Kodaira elliptic fiber with
an Abelian variety of dimension $(r-1)$. When the two ``Kodaira types'' are \emph{not equal} the geometry is subtler.
The existence of a second discrete invariant,\footnote{\ The two notions of ``Kodaira type'' of the singular fiber have quite different nature. While the monodromy class is invariant under continuous deformations (as long as there is no collision between codimension-1 singularities) the type of the characteristic cycle, while a discrete invariant, may jump abruptly under a small deformation of the complex structure.} in addition to the conjugacy class of the local monodromy, explains why two singular fibers with the same monodromy ($\equiv$ physicists' Kodaira type)
may be geometrically and physically distinct.
The main purpose of this paper is to give a physical interpretation of the
Hwang-Oguiso Kodaira type and discuss the physical phenomena controlled by it.

\subparagraph{The Kodaira puzzle.} The reader may be puzzled: the subtlety with two distinct notions of ``Kodaira type'' arises when $r\geq2$. In rank-$1$ the two notions coincide tautologically. Yet it looks that certain singular elliptic fibers come in different ``unfrozen'' vs.\! ``(partially) frozen'' versions. How can be possible?
 
 Of course no geometric subtlety escaped Kodaira. The issue is about physics not geometry. The sentence: ``the $\cn=2$ QFT is described by a rank-1 special geometry over a Coulomb branch of dimension 1'' is less innocent that it may sound. First of all the ``Coulomb branch'' itself is well-defined only up to an equivalence relation which is \emph{strictly weaker} than analytic isomorphism. Second: 
 we typically define our irreducible QFTs by stripping out would-be ``free'' subsectors,
and most ``rank-1 special geometries'' arise as special \emph{sub-}geometries of higher dimensional ones. Often it is not legittimate to ``extract out'' the sub-geometry from
the ambient special geometry because the would-be ``free'' sectors, while decoupled at the level of local physics, globally remain coupled to the ``rank-1 interacting QFT'' by subtle topological effects
 (e.g.\! flux quantization). In this situation only the deformations of the sub-geometry which are induced by global deformations of the ambient special geometry make sense. The puzzle is solved by counting deformations in the proper geometrical set-up which describes the full physics.  
See \S.\,\ref{s:revisited} for details and examples in rank-1.

\subparagraph{Relation to frozen singularities in M- and F-theory.}
The second freezing mechanism is based on multiple Lagrangian fibers:
they were classified by Hwang and Oguiso 
\cite{oguiso2}. This mechanism applies only to fibers whose monodromy has \emph{unstable} type.
A quick comparison of the table of multiple Lagrangian fibers \cite{oguiso2}
with the table of (partially) frozen singularities in M-theory \cite{F2,MM1,MM2,YYY} shows that the two lists are one and the same.
Mathematically this is quite surprising. The two classifications refer to objects which are very far apart in the math space: the table in \cite{F2,MM1,MM2,YYY} arises from the classification of flat bundles on $T^3$ which builds on
 deep facts from the representation theory of compact Lie groups \cite{borel,W4}, while the table \cite{oguiso2} pertains to analytic symplectic geometry. How it happens that they are identical?
 
 While the two stories have no obvious connection in mathematics, physically they are related by a chain of string dualities:
one starts from M-theory with a frozen singularity,  go to the F-theory dual geometry, and probe it with a D3 brane to get a $\cn=2$ QFT on its world-volume. There are a number of subtleties in the physical interpretation of the F-theory geometry from the D3 viewpoint (the ones mentioned above in reference to the Kodaira puzzle). When everything is taken into account the \emph{total} special geometry of the D3 world-volume theory contains a multiple fiber, and its reduction to a rank-1 special \emph{sub}-geometry has a (partially) frozen Kodaira fiber because only the ambient deformations are meaningful, see  \S.\,\ref{s:poiuqw}.
%
%
%

\subparagraph{Implications for the $\cn=2$ SCFT classification program.}
The discussion in this paper has an important impact on the program of classifying
the $\cn=2$ SCFT using special geometry \cite{M1,M2,M3,M4,M5,M6,C1,C2}.
In facts it radically changes the rules of the game, see \S.\,\ref{s:classs}. 
In the last part of the paper we give a rough sketch of the modified classification program in the baby case $r=1$.

\subparagraph{Organization of the note.} Section 2 is meant for the pedantic reader: it spells out the basic definitions of what we mean by ``section'' and ``no-section'' special geometries.
Section 3 is purely physical, even heuristic,
 and aims to introduce the Hwang-Oguiso (HO) invariant for \emph{simple stable} fibers
 from first physical principles. We do this using two alternative languages: line operators (\textbf{Definition 1}) versus partition functions (\textbf{Definition 2}). We also discuss some physical consequences of the HO invariant not being trivial. Section 4 is purely geometric,
 and aims to sketch the structure of the singular fibers in codimension-$1$ following \cite{oguiso1,oguiso2,oguiso3,jap1,jap2,sawon}. Multiple fibers are introduced and their local models and deformation spaces constructed.
 In section \ref{s:general} we return to physics and discuss the physical interpretation 
 of the ``no-section'' special geometries, construct the duality chains which relate geometries with multiple fibers to 
frozen singularities in M- and F-theory. In section 6 we discuss how the HO invariants
\emph{redefine} the geometric classification program of
 $\cn=2$ SCFTs. We apply these ideas to rank-1 with the aim of solving the Kodaira puzzle.

\section{Preliminaries: ``section'' vs.\! ``no-section'' geometries}\label{s:preliminary}

The sole purpose of this section is to specify our basic definitions in order not to confuse the reader with similar sounding but very distinct notions. Readers who think we are too pedantic may jump ahead.

\medskip

Informally speaking, by ``special geometry'' we mean the geometry of the total space of a holomorphic 
integrable system, i.e.\! a holomorphic fibration of complex manifolds
\be
\pi\colon\mathscr{X}\to\mathscr{C}
\ee 
where $\mathscr{X}$ carries a holomorphic symplectic $(2,0)$-form $\Omega$ while the
fibers $\mathscr{X}_u$ ($u\in\mathscr{C}$) are Lagrangian submanifolds: $\Omega|_{\mathscr{X}_u}=0$. There are several distinct variants of this notion, \emph{all} of them play important
 roles in the description of \emph{some} supersymmetric quantum systems.\footnote{\ The physical interpretation of the various notions of special geometry will be discussed elsewhere.} To name just a few alternative definitions:
 \begin{itemize}
\item[\textit{(a)}] one may work in the algebraic category or in the complex analytic one;
\item[\textit{(b)}] one may require $\mathscr{C}$ to be \textit{(i)} compact, or \textit{(ii)} quasi-projective, or \textit{(iii)}
affine, or just \textit{(iv)} a local domain in $\C^r$;
\item[\textit{(c)}] we may or not insists that the smooth loci of the fibers should be algebraic \emph{groups,} in which case we may further require the smooth fibers to be \textit{(i)} proper, or \textit{(ii)} anti-affine, or just \textit{(iii)} Abelian algebraic groups. When the group structure is taken as part of the definition, each fiber $\mathscr{X}_u$ contains a preferred smooth point $0_u$, namely the neutral element of the group law,  $\pi$ has a canonical zero-section 
$s\colon u\mapsto 0_u$, and the base $\mathscr{C}$ is identified with a Lagrangian submanifold of $\mathscr{X}$;
\item[\textit{(d)}] when the generic fibers are proper, we may or may not require them to carry a polarization,
and when we ask for one we may or may not demand it to be principal;
\item[\textit{(e)}] the special geometry may or may not be scale invariant, meaning that there is or not a complete holomorphic vector field $\ce$
(the Euler vector) whose exponential map acts by automorphisms of all structures implied by the particular notion of special geometry while $\mathscr{L}_\ce\mspace{2mu}\Omega=\Omega$.
\end{itemize}

The integrable system of the \emph{usual} 
 Seiberg-Witten geometry \cite{SW1,SW2} of a UV-complete 4d $\cn=2$ QFT
  has an affine base $\mathscr{C}$ (the Coulomb branch) and  polarized
 proper fibers.
The fiber is then a polarized complex torus of dimension $r\equiv\dim_\C \mathscr{C}$ whose physical interpretation is most conveniently described
 by compactifing the QFT down to 3d on a very large circle. By dualizing the 3d vector to a scalar, at low-energy we get a
 8-SUSY $\sigma$-model with target-space a hyperK\"ahler manifold\footnote{\ The hyperK\"ahler metric on $\mathscr{X}$ is the $c$-map of the special K\"ahler metric on $\mathscr{C}$ in the sense of \cite{cmap}. However we shall not be interested in the metric aspects of the geometry.} $\mathscr{X}$ fibered over the 4d Coulomb
 branch $\mathscr{C}$ with generic  fiber  the electric/magnetic holonomy group along $S^1$ polarized by the Dirac pairing.
 The generic smooth fiber of $\pi$ is then automatically a compact group, hence a polarized Abelian variety, and $\pi$ has a zero-section $s$.
 When the QFT is a SCFT the special geometry is, in addition, scale invariant.

 The first notion of ``special geometry'' we shall use in this note is:
 
 \begin{defn}\label{1def} A ``section'' special geometry is a holomorphic fibration $\pi\colon\mathscr{X}\to \mathscr{C}$
 with $\mathscr{X}$ holomorphic symplectic and Lagrangian fibers, whose generic (smooth) fibers are polarized Abelian varieties, and we have a Lagrangian zero-section $s\colon\mathscr{C}\to \mathscr{X}$.
To keep the discussion simple we also assume the polarization to be principal and the geometry to be scale invariant, although these two conditions may be lifted.
 \end{defn} 

The Seiberg-Witten geometry of a typical $\cn=2$ SCFT is a ``section'' special geometry in this sense. The considerations in this paper will be local on the Coulomb branch $\mathscr{C}$, so we do not specify any particular condition on the base.
In practice we work with $\mathscr{C}\equiv \C^r$ ($r$ being the \emph{rank}). 
 We stress that the total space $\mathscr{X}$ is assumed to be smooth.\footnote{\ This is a very mild assumption since typically we can blow up the singularities.}
 
 The second definition is:
  \begin{defn} 
  A special geometry is called ``non-section'' iff all conditions in \textbf{Definition \ref{1def}}
  hold except that there is no section. 
    \end{defn}
    
  \begin{rem}  It follows from our previous considerations that {a ``non-section'' special geometry cannot have the
    physical interpretation of being  the (standard sense) Seiberg-Witten geometry of some $\cn=2$ QFT.}
    Nevertheless ``no-section'' geometries \emph{do have} physical applications/interpretations in supersymmetric QFTs as we explain in section \ref{s:general}.
    \end{rem}
    \begin{rem}  
The singular fibers are necessarily \emph{simple} in presence of a section, while they are allowed to be \emph{multiple} when there is no section. See \S.\,\ref{s:mult}.
\end{rem}

\section{``Section'' special geometry: Physical heuristics}\label{s:heuristics}

In this heuristic section we provide physical intuition for the HO invariant 
in the case of a ``section'' special geometry.
The HO invariant will be introduced from first physical principles.
It turns out (cf. \S.\,\ref{geometry})
that the only \emph{simple} singular fibers which may have a non-trivial HO invariant are the  
\emph{stable} ones, so our heuristic discussion will focus on this case.
The heuristic arguments of this section will exactly match the subtle math theorems of refs.\cite{oguiso1,oguiso2,oguiso3}. In retrospect 
they may be regarded as elegant proofs of those theorems.

\subsection{The discriminant locus}
When the $\cn=2$ QFT is interacting, there is a locus $\mathscr{D}\subset \mathscr{C}$, of pure codimension-1 \cite{jap1}, called the \emph{discriminant,} which parametrizes the vacua where
the physics gets more interesting. Geometrically $\mathscr{D}$ is the (reduced) hypersurface in $\mathscr{C}$ where the fibers are singular:
\be
\mathscr{D}\equiv\{u\in \mathscr{C}\colon \mathscr{X}_u\ \text{singular}\}\equiv \sum_\alpha D_\alpha\quad \text{(as an effective divisor)}.
\ee
We focus on one irreducible component $D_\alpha$ of the discriminant, and fix a \emph{general} point
$u\in D_\alpha$. Going around $D_\alpha$
the periods $(a^i, a^D_j)$ of the Seiberg-Witten differential\footnote{\ In our set-up the Seiberg-Witten differential is just the symplectic dual of the Euler vector $\lambda\equiv \iota_\ce\Omega$.} $\lambda$ get rotated by the local monodromy $\varrho_\alpha\in Sp(2r,\Z)$ with\footnote{\ A codimension-1 locus with trivial local monodromy is not part of the discriminant, i.e.\! the fiber over its generic point is automatically smooth.}
 $\varrho_\alpha\neq1$.
The conjugacy class $[\varrho_\alpha]\subset Sp(2r,\Z)$ of $\varrho_\alpha$ is the basic invariant of $D_\alpha$. $[\varrho_\alpha]$ is independent of the general point $u\in D_\alpha$. We say that the generic fiber over $D_\alpha$
(and $D_\alpha$ itself) is \emph{stable}
if the monodromy $\varrho_\alpha$ is unipotent:
\be
(\varrho_\alpha-1)^2=0.
\ee

As we approach a vacuum $u\in D_\alpha$ some new degrees of freedom become massless: they describe light
BPS  particles whose dynamics is governed by an 
$\cn=2$ effective IR theory
which is either a SCFT in its own right or just IR-free \cite{M2}. Counting dimensions,
we see that the effective IR theory is a rank-1 $\cn=2$ QFT (plus IR-free photons which are massless everywhere in the Coulomb branch). The effective IR theory at the general point $u$ is largely determined by $[\varrho_\alpha]$.
In ``section'' special geometry we have the following dictionary between algebraic properties
of $\varrho_\alpha$ and physical properties of the effective QFT along $D_\alpha$:

\begin{footnotesize}$$
\begin{tabular}{r l|c||c|c|c}\hline\hline
 \multicolumn{3}{c||}{local monodromy $\varrho_\alpha$} &   \multicolumn{2}{c|}{physical properties} & IR effective theory\\\hline
stable: &$\mathrm{rank}(\varrho_\alpha-1)=1$ & non-semisimple & IR free & non UV complete & ?\\\hline
&& non-semisimple & IR free & non UV complete & $SU(2)$ with matter\\
unstable: &$\mathrm{rank}(\varrho_\alpha-1)=2$ & & &  &\\
&& semisimple & SCFT & UV complete & known list of rank-1 SCFT\\\hline\hline
\end{tabular}
$$
\end{footnotesize} 

Some reader may wonder why we put  a question mark in the last box of the first row.
The answer looks pretty obvious: the IR-free, UV non-complete, effective theory is SQED coupled
 to a number of massless
hypermultiplets of charges $q_i$, such that the $\beta$-function coefficient $\sum_i q_i^2$
is equal to $\ell$, the greater positive integer such that $\varrho_\alpha=1\bmod\ell$.

The point is that while this statement may be vaguely correct,
 one should be careful with its precise meaning. When the general fiber over the discriminant component $D_\alpha$
is  \emph{simple} and \emph{unstable} everything is as expected: as we approach the general vacuum $u\in D_\alpha$ the IR effective theory
   asymptotically factorizes in two decoupled subsectors, namely
the $\cn=2$ theory in the last column of the table and  $(r-1)$ massless $\cn=2$ vector multiplets.
By this we mean that,  as we come close to $u$, the partition function of the effective IR theory gets  asymptotically equal  to the 
product of the partition functions of the two subsectors. 
The geometric origin/meaning of this factorization will be explained in \S.\,\ref{geometry}.
For \emph{stable} singular fibers the story is subtler both geometrically and physically as we are going to explain.

\subsection{Physics at a simple stable singularity}

We describe the physical side of the simple stable fiber story in two ways:
in the language of 't Hooft  line operators and in the language of effective Lagrangians and partition functions.
While the second language may be simpler and physically more natural,\footnote{\ The reader may prefer to jump directly to \S.\,\ref{s:2language}.} it is the first language which is more directly
related to the geometric description of the singular fiber.    
\medskip

We stress that the stable situation is characterized by 
the condition $\mathrm{rank}(\varrho_\alpha-1)=1$. 
Since the image of $\varrho_\alpha-1$ is the sublattice of electro-magnetic charges of the BPS particles which get massless as we approach $D_\alpha$, in the stable case \emph{all} light states carry a charge which is an integral multiple of
a fixed charge $q_0$,
\be
\mathrm{im}(\varrho_\alpha-1)=\Z\,q_0.
\ee
We call $q_0$ the (unit) \emph{electric} charge by a redefinition of the duality frame. In particular all light states are mutually
local, are coupled to a single massless photon $A$, and no light monopole or dyon is present. It is easy to check that the light electrically charged hypermultiplets produce the correct $\beta$-function. The effective IR theory \emph{``looks like''} SQED coupled to hypers in the sense that it has
the same spectrum of light BPS \emph{particles.} However this does not imply that 
 the effective IR theory\footnote{\ We take as part of the definition of effective IR theory the truncation to interactions given by an operator of dimension $\leq 4$.} in the general vacuum $u\in D_\alpha$ has the  light BPS \emph{objects} of SQED nor that it factorizes in SQED and decoupled massless vector multiplets.
There are other possibilities;
for their geometric origin see \S.\,\ref{geometry}.

\subsubsection{The language of 't Hooft lines}\label{kkii8888}
While \emph{dynamical} light BPS monopoles are not present in the spectrum, we may mimic them in various ways.
One way is to switch on a topologically non-trivial background for the gauge fields: this is the strategy we shall use in the following subsection. A second
 possibility is to
insert a BPS 't Hooft line operator $H_m^\textsc{ir}(\zeta)$, carrying $m$ units of the effective IR theory magnetic charge,
which we wrap on a circle $S^1_R$ of large radius $R$ with the fixed magnetic holonomy (\textsc{fmh})
periodic boundary condition. The twistor
 parameter $\zeta\in\mathbb{P}^1$ in the symbol $H_m^\textsc{ir}(\zeta)$ specifies the set of 4 supercharges (out of 8) which leave invariant the $\tfrac{1}{2}$-BPS
line operator $H_m^\textsc{ir}(\zeta)$.
The expectation value of this line operator is (by definition) 
\be\label{kjuqwer}
\big\langle H_m^\textsc{ir}(\zeta)\big\rangle_{u,S^1_R}^\textsc{fmh}= \mathcal{X}_m(\zeta,R,u,a)^\text{eff}\Big|_{a\to 0},
\ee   
where the \textsc{rhs} is the ($\C^\times$-valued) hyperK\"ahler function of magnetic charge $m$
defined by GMN \cite{gaiotto}, 
as computed in the effective IR theory,
while $a$ is the IR electric period which is a well-defined global complex coordinate in the Coulomb branch
\emph{germ}\footnote{\ Non UV complete $\cn=2$ models have a Coulomb branch only in the \emph{germ sense.}} of the effective theory. The  (germ of the)
discriminant at $u\in D_\alpha$ is $a=0$. 
\medskip

For the sake of comparison, we recall the situation in $\cn=2$ SQED. For large $R$ one has 
\be\label{ordinary}
\lim_{a\to 0}\mathcal{X}_m(\zeta,R,a)^\textsc{sqed}=\exp[o(R)]\quad\text{for all }m\in\Z,\ \zeta\in\mathbb{P}^1,\ \text{and }R\gg1,
\ee
in 't Hooft's language \cite{H1,H2,H3}: \textit{in ``ordinary'' SQED \emph{all} BPS 't Hooft operators are \emph{light}.}
\medskip

Our situation, eq.\eqref{kjuqwer} is a bit more subtle. The magnetic charge $m$ of the low energy theory is \emph{not} unambiguously
defined: the condition that its Dirac pairing with the
light electric charge $q_0$ is $1$ determines $m$ only up to a linear combination of the $2r-1$
charges which are mutually local with respect to $q_0$.
In other words,   
 to correctly compare the low-energy physics along a simple stable discriminant $D_\alpha$ with ``ordinary'' SQED with the same
light BPS spectrum, eq.\eqref{ordinary}, we have to determine which operator line of the original theory becomes
 the would-be SQED 
't Hooft line as we approach $u$.
This leads to the following\footnote{\ Recall that we are assuming a principal polarization!}:

\begin{defn}\label{ju761234} $D_\alpha\subset\mathscr{C}$ an irreducible component of the discriminant with simple generic fiber and \emph{unipotent} local monodromy $\varrho_\alpha$;
 $u\in D_\alpha$ a \emph{general} point. The \emph{HO invariant} 
 \be
 \mu(u)\in\mathbb{N}\cup\infty
 \ee
  is the smallest
positive integer such that  the effective IR theory along $D_\alpha$ 
admits a twistorial family of  $\tfrac{1}{2}$-BPS 't Hooft lines of magnetic charge
$\mu(u)$, written $H_{\mu(u)}^\textsc{ir}(\zeta)$, which are \emph{light}
in the SUSY vacuum $u\in D_\alpha\subset \mathscr{C}$ for all values of the twistor parameter $\zeta$, that is,
\be
\log \big\langle H_{\mu(u)}^\textsc{ir}(\zeta)\big\rangle_{u,S^1_R}^\textsc{fmh}\equiv\log\mathcal{X}_{\mu(u)}(\zeta,R,u)^\text{eff}=o(R)\qquad \text{as }R\to\infty.
\ee  If there is no \emph{light} BPS  't Hooft line
of any magnetic charge we set $\mu(u)=\infty$. When $\mu(u)=1$ (light lines of all charges)
we say that the HO invariant of the vacuum $u$ is \emph{trivial}.
\end{defn}
\noindent The magnetic charges of light effective 't Hooft lines form the sublattice
$\Z\,\mu(u)\subset \Z$.

\subparagraph{Explicit formulae for rank $r=2$.}
Let $(a^1,a^2)$ be a set of special coordinates\footnote{\ Local coordinates with the required properties exist under our assumptions, see \cite{oguiso3}.} in a neighborhood $U$ of $u$
with $a^2=0$ the germ of $D_\alpha$ at $u$. We write $\{A_i,B^j\}$ for the basis of $H_1(\mathscr{X}_u,\Z)$,
symplectic with respect to the principal polarization,
 such that
($i,j,k=1,2$)
\be
a^i=\int_{A_i}\lambda,\qquad \quad a^D_j=\int_{B^j} \lambda=\tau_{jk}\,a^k,
\ee
where the last equality follows from scale invariance. 
The period matrix of the Abelian fiber in the symplectic basis $\{A_i,B^j\}$ is
\be
\tau_{ij}=\frac{\partial^2\cf}{\partial a^i\partial a^j},\qquad i,j=1,2
\ee
where the prepotential $\cf$ has the local form
\be
\cf(a^1,a^2)=\widetilde{\cf}(a^1,a^2)+\frac{\ell}{4\pi i}\,(a^2)^2\log a^2,\qquad\ell\in \mathbb{N}
\ee
with $\widetilde{\cf}(a^1,a^2)$ holomorphic in $U$.
From \textbf{Definition \ref{ju761234}}, the HO invariant $\mu(a^1)$ is the smallest positive integer such that we can find
two integers $(s,t)\in\Z^2$ with 
\be
\lim_{a^2\to 0}\Big(\mathcal{X}_{B^2}(\zeta,R,a^1,a^2)^{\mu(a^1)}
\mathcal{X}_{A_1}(\zeta,R,a^1,a^2)^{s}\mathcal{X}_{B^1}(\zeta,R,a^1,a^2)^{t}\Big)=e^{o(R)}\quad\forall\;\zeta.
\ee 

The logarithm of the \textsc{lhs} is \cite{gaiotto}
\be\label{pioqw1}
\frac{R}{\zeta}\,\cb(a^1,\mu(a^1), s,t)\,a^1+R\,\zeta\; \overline{\cb(a^1,\mu(a^1), s,t)}\;\overline{a}^1+ o(R)
\ee
where
\be
\cb(a^1,\mu(a^1), s,t)\overset{\rm def}{=} 
\mu(a^1)\mspace{2mu}\frac{\partial^2\widetilde{\cf}}{\partial a^2\partial a^1}\bigg|_{a^2=0}+ s+t\,\frac{\partial^2\widetilde{\cf}}{\partial a^1\partial a^1}\bigg|_{a^2=0}.
\ee 

The $O(R)$ term in \eqref{pioqw1} must vanish for all $\zeta\in\mathbb{P}^1$. Since the point $u$ is general in the divisor $a^2=0$, in particular its first coordinate $a^1\neq0$. 
We conclude that in the context of  {\bf Definition \ref{ju761234}} (i.e.\! for $u$ generic point in a simple stable discriminant component) one has:
\begin{quote}
\it  In the rank-$2$ case
$\mu(a^1)$ is the smallest positive integer such that 
\be\label{kkkiqwerttt}
\cb(a^1,\mu(a^1), s,t)=0\quad \text{for some integers $s$, $t$}.
\ee
\end{quote}\vskip-12pt
 Equivalently:\footnote{\ Compare with \textbf{Remark 5.2} of \cite{oguiso3}.}
\begin{quote}\it In rank-$2$  $\mu(a^1)$
is the order of the complex number $\exp[2\pi i \,\partial_{a^2}\partial_{a^1}\widetilde{\cf}(a^1,0)]$
in the cyclic subgroup of $\C^\times$ generated by $\exp[2\pi i \,\partial_{a^1}^2\widetilde{\cf}(a^1,0)]$.
\end{quote}

\subparagraph{Generalization to rank $r$.} Let $a^i$ ($i=1,\dots,r$)
be local special coordinates such that the stable discriminant germ is given by $a^r=0$. We set $u\equiv (a^1,\cdots, a^{r-1},0)$.
Eq.\eqref{pioqw1} gets replaced by
\be
\begin{split}
\lim_{a^r\to  0} &\log \mathcal{X}_{\mu(u)}(\zeta,R, u, a^r)=\\
&= \frac{R}{\zeta}\,\sum_{i=1}^{r-1} \cb(u,\mu(u), s_i, t^j)_i\, a^i+
R\,\zeta\, \sum_{i=1}^{r-1} \overline{\cb(u,\mu(u), s_i, t^j)}_i\,\overline{a}^i+o(R)
\end{split}
\ee 
for the $(r-1)$-component vector $\big(\cb(u,\mu(u), s_1, t^j)_1,\dots,\cb(u,\mu(u), s_{r-1}, t^j)_{r-1}\big)\in\C^{r-1}$
\be
\cb(u,\mu(u), r_i, s^j)_i= \mu(u) \frac{\partial\widetilde{\cf}}{\partial a^i \partial a^r}\bigg|_{a^r=0} + r_i +s^j \frac{\partial^2\widetilde{\cf}}{\partial a^j \partial a^i}\bigg|_{a^r=0},\qquad
(r_i,s^j)\in\Z^{2r-2}.
\ee

The statement that $u\in D_\alpha$ is generic means that the complex numbers $a^i$, $a^D_i$ ($i=1,\dots,r-1$)
are linear independent over $\mathbb{Q}$. Then
\begin{quote}\it
$\mu(u)$ is the order of the element $\exp[2\pi i \partial_{a^i}\partial_{a^r}\widetilde{\cf}]\in (\C^\times)^{r-1}$
in the subgroup $\Z^{r-1}\subset (\C^\times)^{r-1}$ generated by the $(r-1)$ elements 
$\exp[2\pi i \partial_{a^i}\partial_{a_j}\widetilde{\cf}]\in  (\C^\times)^{r-1}$ for $j=1,2,\cdots,r-1$.
\end{quote}

Clearly the integer $\mu(u)$ is an important invariant of the fiber at a general point $u$ on a stable divisor $D_\alpha$,
related to the invariant defined by Hwang and Oguiso \cite{oguiso1,oguiso2,oguiso3}. The precise relation of $\mu(u)$ with their characteristic cycle $\boldsymbol{\Psi}_u$
will be spelled out in the section \ref{geometry}.

\subsubsection{The language of effective Lagrangians and partition functions}\label{s:2language}
The low-energy theory at the general vacuum $u$ in a simple stable divisor $D_\alpha$
 consists of $r$ massless photons coupled to
a number of hypers which are charged only with respect to one particular
 linear combination $A$ of the light gauge fields.
 
 At the global level the gauge field $A$ is not just a vector field, it is a $U(1)$ gauge connection
 on some line bundle $\cl$ over spacetime. To fully specify the low-energy field theory,
 we have to specify the set of gauge topological sectors over which we sum in the
 path integral \cite{reading}. The inequivalent SQED theories are labelled by a non-negative integer $k$.
 In $k$-SQED we sum over line bundles such that $k\mid c_1(\cl)$. 
 Fixing a non-zero $k$ in the partition sum is the second way in which we may mimic the presence of magnetic monopoles
 besides the introduction of 't Hooft lines as we did in the first part of this section.
 We write $Z_k^\text{eff}$ for the partition function of the effective IR theory at $u\in D_\alpha$,
 quantized in a periodic Euclidean box,
 where we restrict $A$ (the field which couples to the light hypers)
 to be a connection on line bundles $\cl$ with $c_1(\cl)$ a multiple of $k$, and no other topological restriction, except
 the ones already implied everywhere in $\mathscr{C}$ by Dirac quantization of the original SCFT (which enforces the principal polarization on the fibers).
 In this language \textbf{Definition \ref{ju761234}} becomes:

 \begin{defn} The \emph{HO invariant} $\mu(u)\in\mathbb{N}\cup\infty$ is the smallest positive integer such that the path integral of the effective IR theory with $k=\mu(u)$ factorizes \emph{asymptotically}
 \be\label{juuyyqwer7}
 Z_k^\text{eff}\leadsto Z_k^\text{SQED} \times Z^\text{free}
 \ee
 in the path integral of $k$-SQED and a partition function for $(r-1)$
 free vector multiplets.
 \end{defn}
 
 Let us explain. 
The kinetic terms at $u$ mix the gauge field $A$ with the other $(r-1)$ massless vectors.
At the level of local physics, we can always diagonalize the vector fields' kinetic terms
undoing the mixing. However, to achieve this diagonalization, we have to redefine the vector fields by taking linear combinations of them 
with \emph{real} coefficients. This operation will diagonalize the vectors' kinetic terms, but not (in general) also the topological $\theta$-terms $\theta^{ab} F_a\tilde F_b/16\pi^2$ which are irrelevant
for the local physics but have observable global effects in non-trivial backgrounds.
Thus \emph{as far as the local physics is concerned} the factorization (asymptotically) holds. In particular the perturbative
amplitudes factorize.
 However the field redefinition is illegitimate at the global level: a linear combination with real coefficients of $U(1)$ connections in general is not a $U(1)$ 
connection since its fluxes are not properly quantized.
So, in general, the factorization \eqref{juuyyqwer7} does not hold  for any $k$ because the two
would-be sub-sectors are still coupled by flux quantization (and also by the $\theta$-terms in a topologically non-trivial background). When the coefficients happen to be integral, and the same linear combination also diagonalizes
the $\theta$-terms, the factorization \eqref{juuyyqwer7} holds in all backgrounds.\footnote{\ \emph{A priori} the statement is valid whenever the field redefinition
matrix  is in $GL(r,\Z)$. In our case the redefinition matrix may be taken to be unipotent (cf.\! eq.\eqref{poiuqwer}) so
 it is automatically in $GL(r,\Z)$ if it has integral coefficients.}

An intermediate situation is when the coefficients are rational. Suppose that 
 the linear combinations which diagonalize both the kinetic terms and the $\theta$-terms are (say for $r=2$)
\be\label{poiuqwer}
A_1^\prime = A_1+\frac{a}{b} A_2,\qquad A_2^\prime = A_2
\ee  
with $A_2^\prime\equiv A_2$ the field coupled to the light hypers (which is automatically a \emph{bona fide} connection).
In eq.\eqref{poiuqwer} $a$, $b$ are coprime integers. Let $\cl$ be the line bundle associated to $A_2$
(so the light hypers are sections of some powers of $\cl$). From eq.\eqref{poiuqwer} we see that
$A_1^\prime$ is a genuine gauge connection iff
$b$ divides $c_1(\cl)$. Therefore we have the factorization \eqref{juuyyqwer7} iff we choose $k$
in the definition of the theory to be a multiple of $b$. In other words: $k\equiv b$ is the minimal 
integer for which we have factorization.

In our set-up $b$ coincides with the HO invariant $\mu(u)$, as it is already clear from the previous subsection and the fact that we can
define only sectors with magnetic charges (i.e.\! fluxes) $c_1(\cl)$ which are multiples of $b$. 
To see the  equality $b\equiv\mu(u)$ more explicitly, let us write the part of the effective Lagrangian in the Coulomb vacuum $u$ which contains the vectors' kinetic and $\theta$-terms
\be
\begin{split}
\frac{1}{16\pi i}\,&\begin{pmatrix}F^+_1 &F^+_2\end{pmatrix}\!\begin{pmatrix}\tau_{11} & \tau_{12}\\
\tau_{12} & \tau_{22}\end{pmatrix}\!\begin{pmatrix}F^+_1\\ F_2^+\end{pmatrix}+\text{h.c.}=\\
&=\frac{1}{16\pi i}\left[\left(\tau_{22}-\frac{\tau^2_{12}}{\tau_{11}}\right)\!(F_2^+)^2+
\tau_{11}\left(F_1^++\frac{\tau_{12}}{\tau_{11}} F_2^+\right)^{\!2}\right]+\text{h.c.},
\end{split}
\ee
where $F_i=dA_i$ are the gauge field-strengths and $F_i^\pm = (F_i\pm i \tilde F_i)/2$
their (anti)selfdual projections. Comparing with \eqref{poiuqwer} we see that the minimal $b$
is the order of $\tau_{12}/\tau_{11}$ in $\C/\Z$ or, more precisely, (given that we may change
$\tau_{11}$ by any integer by shifting the corresponding $\theta$-angle by a multiple of $2\pi$
leaving the partition function unchanged), it is given by the order
of $\exp(2\pi i\mspace{2mu} \tau_{12})$ in the multiplicative group generated by $\exp(2\pi i\mspace{2mu} \tau_{11})$.
But this is precisely the definition of the HO invariant $\mu(u)$, eq.\eqref{kkkiqwerttt}.
The extension to arbitrary rank $r\geq2$ is straightforward.

\subsubsection{HO invariant vs.\! flavor symmetry}

In the vicinity of the vacuum $u\in D_\alpha$ the effective gauge coupling
$g_\text{eff}$ of the Abelian field $A$, coupled to the light BPS states, is
\be
\frac{\theta_\text{eff}}{2\pi}+\frac{4\pi i}{g_\text{eff}^2}= \frac{\ell}{2\pi i}\log a+\text{regular}
\ee
and hence $g_\text{eff}\to 0$ as we approach $u$ (i.e.\! as $a\to 0$).
At zero coupling a gauge symmetry reduces  to a flavor (global) symmetry. Hence, \emph{a priori,}
we get a $U(1)$ flavor symmetry for the residual rank-$(r-1)$ $\cn=2$ QFT which remains after  decoupling the
IR-free light sector. (The residual rank-$(r-1)$ theory is well-defined, at least as an IR effective theory,
see \S.\,\ref{geometry}). 

We may ask: does this $U(1)$ flavor symmetry act non-trivially on the rank-$(r-1)$ residual model?
For instance, if the original rank-$r$ theory was the decoupled product of a rank-$(r-1)$ and
a rank-1 QFT, and $D_\alpha$ was the component of the discriminant $\C^{r-1}\times\{0\}$,
the $U(1)$ would not act on the rank-$(r-1)$ model. However physical intuition says that
the $U(1)$ action on the rank-$(r-1)$ QFT is ``typically''  non-trivial. 
How can we determine whether this is really the case?

A detailed answer requires the special geometric theory of flavor which will be introduced elsewhere \cite{toappear}.
We sketch the physics of the reasoning. In $\cn=2$ supersymmetry, the mass parameters may be seen as
Coulomb branch coordinates of a zero-coupling SUSY gauge theory with the flavor group as gauge group.
On the other hand, the mass parameters enter the meromorphic Seiberg-Witten differential as residues at the poles. 
Hence at $a=\epsilon$, the Seiberg-Witten differential of the rank-$(r-1)$ QFT has poles with residue of order $\epsilon$ if and only if
the group $U(1)$ acts effectively on the rank-$(r-1)$ theory. One checks that a pole with residue $O(\epsilon)$ 
is present if and only if the HO invariant is non-trivial. The previous example of the product of a rank-$(r-1)$
and a rank-$1$ theory makes this conclusion physically intuitive.

\subsubsection{The HO invariant as a Higgs branch obstruction}

In \S.\,2.7 of \cite{M2}  Argyres and Martone distinguish two physical situations around a divisor with $\mathrm{rank}(\varrho_\alpha-1)=1$ (a stable degeneration). While their geometric characterization of the two cases is a bit vague,
we gather that their two different behaviors correspond in our language to, respectively, trivial and non-trivial HO invariant.
Motivated by some physical considerations, they suggest the following (in our language):
\begin{quote}\it
A non-trivial HO invariant is an obstruction to the existence of a Higgs branch for the
rank-$1$ effective IR theory governing the new light degrees of freedom. 
\end{quote}
They also present a number of examples supporting the statement.

This is yet another way of saying that the two sectors does not asymptotically factorize  because they remain coupled through flux quantization. The Higgs phase is characterized by the presence of stable
magnetic flux-tubes which do not spread out \cite{H1,H2,H3}. Suppose the gauge field $A$
of the asymptotic rank-1 theory is in the Higgs phase, so that we have stable magnetic flux-tubes
of unit flux. From eq.\eqref{poiuqwer} we see that this is also a stable magnetic flux tube for the
$A^\prime_1$ gauge field with $\tfrac{a}{b}\bmod1$ units of flux.  So the residual sector, which contains the light gauge field $A^\prime_1$, is not in its Coulomb phase, contrary to the assumptions.

\subsubsection{Wild nature of $\mu(u)$}\label{s:wild}

We have seen that the integer $\mu(u)$ has a direct physical meaning. Indeed, 
$\mu(u)$ controls the ``quantum phase''\footnote{\ In the sense of 't Hooft's theory of phases of 4d quantum gauge models \cite{H1,H2,H3}.} of the low-energy effective
gauge theory in the vacuum $u\in D_\alpha\subset\mathscr{C}$.
Despite this, the invariant behaves quite wildly under infinitesimal deformations of the reduced prepotential $\widetilde{\cf}$.
In  (say) rank-2 $\mu(a^1)<\infty$ requires the three complex numbers
\be
1,\qquad \tau_{11}(a^1)\equiv \frac{\partial^2\widetilde{\cf}}{\partial a^1\partial a^1}\bigg|_{a^2=0}\!,
\qquad \tau_{21}(a^1)\equiv\frac{\partial^2\widetilde{\cf}}{\partial a^2\partial a^1}\bigg|_{a^2=0}
\ee
to be linearly dependent over $\mathbb{Q}$, which is an extraordinarily non-generic case. 
Only for special and quite sharp values of the coefficients in the reduced pre-potential $\widetilde{\cf}$
this number-theoretic condition will be fulfilled. 
When the condition holds, it will be typically destroyed by any deformation, however small.
Consider, for instance, deforming $\widetilde{\cf}\leadsto \widetilde{\cf}+t\, (a^1)^2/2$
with $|t|<\epsilon$. Now we need $1$, $\tau_{11}+t$ and 
$\tau_{21}$ to be linearly dependent over $\mathbb{Q}$:
this is most certainly false for almost all $t$'s in any small disk $|t|<\epsilon$ (unless  $\tau_{21}\in\mathbb{Q}$).

\begin{rem} In rank-$2$ SCFT the HO invariant has the same value at all general points of a simple stable irreducible component of
the discriminant. This follows from the fact that $\C^\times$ acts by complex analytic isomorphisms of the fibers, so
it preserves the HO invariant, while $D_\alpha$ is the closure of an open $\C^\times$-orbit. 
On the contrary, in higher rank (or in a non-conformal $r=2$ model)
the invariant is expected to behave quite unruly as we vary $u$ in $D_\alpha$: a \emph{generic} pre-potential will lead to $\mu(u)=\infty$ almost everywhere in $D_\alpha$,
except for an everywhere dense zero-measure subset, see the examples in \S.\,5 of \cite{oguiso3}.  
\end{rem}

\begin{rem} The case $\mu(u)=\infty$ has a characterization in terms of algebraic groups. The connected component of the smooth locus of the fiber over $u$ is an \emph{anti-affine algebraic group} if and only if $\mu(u)=\infty$
i.e.\! if there is no light magnetic line. The bearing of this fact will be explored elsewhere \cite{toappear}.
\end{rem}

\subsubsection{The fiber freezing mechanism}

To preserve the value of $\mu(a^1)$ the infinitesimal deformation should have quite a specific form. Usually there will be no deformation
with this ``magic'' property. Thus, when physical consistency (which coincides with \emph{geometric} consistency) requires the  HO invariant $\mu$ to have
a specific \emph{finite} value, the singular fiber over $u$ gets rigidified, or \emph{frozen} as people like to say,
and very few (or no) deformations survive. This is a subtle geometric mechanism to freeze exceptional fiber
which cannot be detected by the local monodromy around it.

The case of HO invariant 1 looks different. 
A stable fiber with $\mu(u)=1$ shares many of the geometric properties of the
unstable case (we shall make this precise in \S.\,\ref{geometry}), and just as we do not expect any
freezing for the \emph{simple} unstable fibers, we neither expect it for stable fibers of invariant 1.
There are several physical arguments which point in this direction.
First: there is the evidence from the known examples: in ref.\cite{M2} the term
\emph{frozen} appears to be used interchangeably with their condition which we interpret as non-trivial
HO invariant. Second: from the QFT viewpoint trivial HO invariant is equivalent
to the (asymptotic) factorization of the $k=1$ partition function, eq.\eqref{juuyyqwer7}. All perturbations of the effective QFT which preserve factorization of the partition function should correspond to deformations of the geometry which keep the 
HO invariant equal 1. In particular, this applies to deformations which act independently on the two sectors.  
Third: the presence of a discrete symmetry typically forces the invariant to be 1. 
For instance, when the pre-potential $\widetilde{\cf}(u,a)$ is
invariant under $a\leftrightarrow-a$, we have automatically
HO invariant 1.
 All deformations which preserve the discrete symmetry also
preserve $\mu(u)=1$.  Fourth: when a Higgs branch is present (and the electric charge of the hypermultiplets is 1)
all deformations which do not destroy the Higgs branch should preserve $\mu(u)=1$.

Geometrically a trivial HO invariant means that the fiber is the product of a 
rank-$(r-1)$ Abelian variety and stable elliptic singular fibers. The deformations of the singularity coincide with the deformation of the stable elliptic fiber, i.e.\! the deformation space is the one naively associated to the conjugacy class of the local monodromy.

\section{Interlude: Geometry}\label{geometry}

In this section we sketch the geometric structure of the singular fibers in codimension-1 of
the holomorphic Lagrangian fibration 
\be
\pi\colon\mathscr{X}\to \mathscr{C},\qquad \dim_\C\mathscr{X}=2\,\dim_\C\mathscr{C}=2r,
\ee 
following
\cite{oguiso1,oguiso2,oguiso3,jap1,jap2,sawon}. Our discussion will be rather informal, aiming to convey the
general picture of the geometry. Readers looking for precise theorems and rigorous proofs are referred
to the original math literature \cite{oguiso1,oguiso2,oguiso3,jap1,jap2,sawon}.
We assume the polarization to be principal, although most results stand without this condition.
\medskip

%
%
%

The singular fibers in codimension-1 are modelled\footnote{\ For the precise meaning of the term ``modelled'' see 
\textbf{Definition 4.1} in \cite{oguiso2}.} on the elliptic Kodaira fibers \cite{koda1,koda2,koda3,bhm} which are
the rank-$1$ case. We start by reviewing that story, and then look for its possible generalizations in rank $r$.

\subsection{Review of the rank-1 case}\label{s:rank1} 
In rank-1 we have a holomorphic fibration 
$\pi\colon S\to B$ of a smooth complex surface $S$ over a (non-compact) curve $B$ whose generic (smooth)
 fibers are elliptic curves. We assume 
the fibration to be relatively \emph{mimimal} i.e.\! no curve with self-intersection $-1$ is contained in a fiber.
The considerations are local in $B$, and we may take $B\equiv\Delta$,
a small disk centered at a point with an exceptional non-smooth fiber.
This is the classical situation studied by Kodaira in his seminal papers \cite{koda1,koda2,koda3} where
he classifies the 
allowed singular fibers into \emph{types}. 
One may give several different mathematical interpretations of Kodaira's classification: most of them were already spelled out  in the original papers \cite{koda1,koda2,koda3}. We briefly outline some of them for later comparison with the corresponding situations in higher dimension.

\subparagraph{Intersection theory viewpoint.}
The exceptional fiber $S_u$ is a divisor of the form
\be
S_u\equiv \pi^{-1}(u)=\sum_i m_i\mspace{2mu} \Theta_i
\ee 
where $\Theta_i\subset S$ are rational curves (copies of the Riemann sphere $\mathbb{P}^1$),
$m_i$ positive integers (\emph{multiplicities}), and $S_u^2=S_u\cdot S_{u^\prime}\equiv 0$ ($u^\prime\neq u$) since all fibers are homologous. 
The divisor $S_u$ has connected support.
The multiplicity of the fiber is $m\equiv\gcd(m_i)$: the fiber is \emph{simple} when $m=1$.
We focus on the simple fibers. The fiber is necessary simple when there is a local section $s$; indeed
\be
1=S_u\!\cdot\! s\equiv \sum_i m_i\mspace{2mu} (\Theta_i\!\cdot\! s),
\ee
which means that the zero-section $s$ meets a single component
$\Theta_0$ with $m_0=1$.

The \emph{type} of the simple fiber $S_u$ is determined by the multiplicities $\{m_i\}$ and the way 
  the various irreducible components $\Theta_i$ cross each other.
By Zariski's lemma \cite{bhm} the  intersection form $\Theta_i\!\cdot\! \Theta_j\equiv-C_{ij}$ is a negative semi-definite, symmetric, integral matrix, with diagonal entries\footnote{\! Self-crossing number $-1$ is ruled out by the minimality assumption.} $0$ or $-2$ and non-negative off-diagonal ones. Moreover $C_{ij}$ has a zero-eigenvector with positive integral entries, $C_{ij}m_j=0$. Indeed
\be
0\equiv\Theta_i\!\cdot\! S_{u^\prime}=\Theta_i\!\cdot\! S_u=\sum_j m_j\,\Theta_i\!\cdot\!\Theta_j= - C_{ij}\,m_j,\qquad
u^\prime\neq u.
\ee
Then $C_{ij}$ must be the direct sum of Cartan matrices of simply-laced affine Kac-Moody Lie algebras and zero-matrices. Since $S_u$ is connected, $C_{ij}$ is zero iff the fiber is irreducible $S_u=\Theta$. When the fiber is reducible $-\Theta_i\!\cdot\!\Theta_j$ is 
 the Cartan matrix of a simple, simply-laced, affine Lie algebra of type $\widetilde{A}_{n-1}$, $\widetilde{D}_{4+n}$, or $\widetilde{E}_s$ ($s=6,7,8$). In the irreducible case the fiber is a singular rational curve of virtual genus 1 (a singular cubic), and we have two possibilities:
either the rational curve $\Theta$ has a simple node (fiber of Kodaira type $\mathbf{I}_1$) or a cusp (Kodaira type $\mathbf{II}$).
The intersection matrices of types $\widetilde{A}_1$ (resp.\!  $\widetilde{A}_2$) is realized by two different configurations of $\mathbb{P}^1$'s:
either two (resp.\! three) rational curves tangent (resp.\! crossing transversely) at a single point
which is Kodaira type $\mathbf{III}$ (resp.\! $\mathbf{IV}$), or two (resp.\! three) rational curves crossing transversely
at distinct points which is type $\mathbf{I}_2$ (resp.\! type $\mathbf{I}_3$). All other $\widetilde{A}\widetilde{D}\widetilde{E}$ intersection matrices are realized only by the
configuration of rational curves which is dual to the affine Dynkin graph $\widetilde{\Gamma}_\mathfrak{g}$ of the corresponding Kac-Moody algebra $\widetilde{\mathfrak{g}}$: a simple root  $\alpha_i\in\widetilde{\Gamma}$ represents a rational component $\Theta_i$, and two distinct components, 
$\Theta_i$ and $\Theta_j$ ($i\neq j$), cross transversely at a number of distinct points equal to $-C_{ij}$, each curve having intersection number $-2$ with itself. The equation $C_{ij}\mspace{1mu}m_j=0$ identifies the multiplicity $m_i$ of $\Theta_i$ with the Coxeter label of the simple root $\alpha_i$. 
The dual configurations to the affine Dynkin graphs of types $\widetilde{A}_{n-1}$, $\widetilde{D}_{4+n}$, $\widetilde{E}_6$,
$\widetilde{E}_7$, and $\widetilde{E}_8$
are called, respectively, exceptional fibers of Kodaira type $\mathbf{I}_n$, $\mathbf{I}^*_n$, $\mathbf{IV}^*$, $\mathbf{III}^*$, and  $\mathbf{II}^*$. 
Here $n\geq0$: $\mathbf{I}_0$ stands for a smooth elliptic curve, and $\mathbf{I}_1$ for the rational curve with a simple node.

\subparagraph{Monodromy perspective.} Alternatively we can classify singular elliptic fibers using the
period map (called the \emph{functional invariant} by Kodaira \cite{koda2}) and the associated monodromy representation
(his \emph{homological invariant} \cite{koda2}).
The possible (simple) fibers are naturally in one-to-one correspondence with the conjugacy classes
of local monodromies around a point which are consistent with the strong monodromy theorem \cite{vhs2,periods} and the
nilpotent orbit theorem \cite{smit} for variations of Hodge structure \cite{vhs1,vhs2,vhs3,vhs4,vhs5,vhs6,periods,smit}.

The allowed local monodromies in rank-1 are the elements $\rho\in SL(2,\Z)$ which
are quasi-unipotent and satisfy the non-negativity condition of the
nilpotent orbit theorem\footnote{\ $\Omega$ is the symplectic form preserved by $SL(2,\Z)\equiv Sp(1,\Z)$.}
\be
(\rho^m-1)^2=0\quad\text{and}\quad \mathrm{tr}\big(\Omega(\rho^m-1)\big)\geq0,\quad\text{for some $m\in\mathbb{N}$.}
\ee 
The explicit correspondence between Kodaira types and conjugacy classes \cite{koda1}
is recalled in table \ref{table} for the fibers with semisimple monodromy and in table
\ref{table2} for the ones with non-semisimple monodromy.

\begin{table}
\caption{\label{table} Semisimple Kodaira fibers}\vskip9pt

{\begin{footnotesize}\renewcommand{\arraystretch}{1.5}
\begin{tabular}{p{2.2cm}p{1.25cm}p{1.5cm}p{1.35cm}p{1.35cm}p{1.35cm}p{1.35cm}p{1.4cm}p{1cm}}
\hline\noalign{\smallskip}
Kodaira type & $\mathbf{I}_0$ & $\mathbf{I}_0^*$ & $\mathbf{II}$ & $\mathbf{II}^*$ & $\mathbf{III}$ & $\mathbf{III}^*$ & $\mathbf{IV}$ & $\mathbf{IV}^*$\\
\noalign{\smallskip}\hline\noalign{\smallskip}
$m$ & $1$ & $2$ & $6$ & $6$ & $4$ & $4$ & $3$ & $3$\\
monodromy & $\left[\begin{smallmatrix}1 & 0\\
0 & 1\end{smallmatrix}\right]$ & $\left[\begin{smallmatrix}-1 & 0\\
0 & -1\end{smallmatrix}\right]$ & $\left[\begin{smallmatrix}1 & 1\\
-1 & 0\end{smallmatrix}\right]$ &$\left[\begin{smallmatrix}0 & -1\\
1 & 1\end{smallmatrix}\right]$ & $\left[\begin{smallmatrix}0 & 1\\
-1 & 0\end{smallmatrix}\right]$ & $\left[\begin{smallmatrix}0 & -1\\
1 & 0\end{smallmatrix}\right]$ & $\left[\begin{smallmatrix}0 & 1\\
-1 & -1\end{smallmatrix}\right]$ &$\left[\begin{smallmatrix}-1 & -1\\
1 & 0\end{smallmatrix}\right]$\\
Euler no. & $0$ & $6$ & $2$ & $10$ & $3$ & $9$ & $4$ & $8$\\ 
Lie algebra $\mathfrak{g}$ & $-$ & $\mathfrak{so}(8)$ & $-$ & $E_8$ & $\mathfrak{su}(2)$ & $E_7$ & $\mathfrak{su}(3)$ & $E_8$\\
\noalign{\smallskip}\hline\noalign{\smallskip}
\end{tabular}\end{footnotesize}}
\end{table}\begin{table}
\caption{\label{table2} Non-semisimple Kodaira fibers. Here $n\geq1$}\vskip9pt
\centering
{\begin{small}\renewcommand{\arraystretch}{1.5}
\begin{tabular}{p{3cm}p{1.5cm}p{1.8cm}}
\hline\noalign{\smallskip}
Kodaira type & $\mathbf{I}_n$ & $\mathbf{I}_n^*$ \\
\noalign{\smallskip}\hline\noalign{\smallskip}
monodromy & $\left[\begin{smallmatrix}1 & n\\
0 & 1\end{smallmatrix}\right]$ & $\left[\begin{smallmatrix}-1 & -n\\
0 & -1\end{smallmatrix}\right]$\\
Euler number & $b$ & $b+6$ \\ 
Lie algebra $\mathfrak{g}$ & $\mathfrak{su}(n)$ & $\mathfrak{so}(2n+8)$ \\
\noalign{\smallskip}\hline\noalign{\smallskip}
\end{tabular}\end{small}}
\end{table}

\subparagraph{Group theory viewpoint.} We are assuming the fibration $\pi$ to have a section $s$.
The \emph{generic} fiber $S_{u^\prime}$ is an elliptic curve with a marked zero point,
$s(u^\prime)\subset S_{u^\prime}$, hence (in particular) is 
an algebraic Abelian group of complex dimension 1 \cite{MB}.
A simple singular fiber $S_u$ is not a commutative group, but its subset 
 $S^\sharp_u\subset S_u$ of smooth points inherits a commutative group structure from the nearby smooth fibers.
 Thus $S^\sharp_u$ is a non-proper algebraic Abelian group of complex dimension 1, hence the extension by a finite
 commutative group of 
either $\C^\times$ (multiplicative case) or $\C$ (additive case). One can show that the finite group is isomorphic to the center
of the simply-connected Lie group with the Lie algebra associated to the exceptional fiber (cf. last row of tables \ref{table},\,\ref{table2}).
 The classification of fibers can be recovered by looking to the blow-ups and blow-downs
 of components of such a group-fiber which are needed to make the total space $S$ locally smooth around the
 singular fiber.
 See \cite{koda1,koda2,koda3} for details.
 
 \begin{rem} In complex dimension 1 the multiple fibers are necessary of type $\mathbf{I}_n$
 and they cannot appear in Lagrangian fibrations. In higher rank the story
 will be radically different.
 \end{rem}

\subsection{Higher rank: simple fibers}\label{s:simplefib}

Our task is to generalize the Kodaira theory to higher rank. We have the embarrassment of riches:
which one of the several interpretations of the rank-1 classification are we supposed to generalize to higher dimension?

Even if we focus on the intersection viewpoint, we still remain with a two-fold embarrassment:
have we to look for an intersection pattern of the $r$-dimensional irreducible components of the exceptional fiber, 
or have we to look for a maximal connected cycle of rational curves in the fibers? In rank-1
the two objects are tautologically identical, but they are obviously quite different in higher $r$.
We shall see in this subsection that both strategies are possible, both lead to a classification in terms of Kodaira types,\footnote{\ The author of \cite{jap1,jap2} makes the first choice, the authors of \cite{oguiso1,oguiso2,oguiso3} the second one.}
but the two viewpoints may lead to two \emph{different} Kodaira types for a given fiber $\mathscr{X}_u$.
The ``discrepancy'' between the two is what we call the ``HO invariant'' (when the fiber is simple): it is a measure of how far from the naive picture the actual geometry is.

To generalize the intersection theoretic approach we need to define the multiplicities 
and the intersection numbers of the relevant objects.

\subsubsection{Multiplicities}\label{s:mult}
In a Lagrangian fibration
the inverse image of the irreducible discriminant divisor $D_\alpha\subset \mathscr{C}$ has the form
\be
\pi^{-1}(D_\alpha)=\sum_a m_{\alpha,a}\, E_{\alpha,a}
\ee
 for certain positive
integers $m_{\alpha,a}$ and irreducible effective divisors $E_{\alpha,a}\subset \mathscr{X}$. 
Let $u\in D_\alpha$ be a \emph{generic} point,
and $\{F_{u,i}\}_{i\in I_u}$ be the (finite) set of irreducible components of the singular
 fiber $\mathscr{X}_u\equiv \pi^{-1}(u)$.

\begin{defn}\label{poq12zmm}
The \emph{multiplicity} $m_{u,i}$ of the fiber component $F_{u,i}$ is 
the multiplicity $m_{u,i}\equiv m_{u,a(i)}$ of the irreducible divisor $E_{\alpha,a(i)}$
which contains it. We identify the fiber $\mathscr{X}_u$
with the cycle 
\be\label{iu}
\mathscr{X}_u\leadsto \sum_{i\in I_u} m_{u,i}\, F_{u,i}.
\ee
 We call $m_u\equiv \gcd(m_{u,i})$ the multiplicity of the fiber at $u\in D_\alpha$.
The fiber $\mathscr{X}_u$ is \emph{simple} iff $m_u=1$, otherwise it is \emph{multiple.}
\end{defn}
\noindent When our Lagrangian fibration has a section $\mathscr{C}$ (whose image in a regular fiber
 is the role of neutral element
of the Abelian fiber), for $u$ a generic point in the discriminant we have
\be
1=\mathscr{C}\cdot\mathscr{X}_u= m_u\,\sum_i \frac{m_{u,i}}{m_u}\; \mathscr{C}\cdot F_{u,i}
\ee
which implies that $\mathscr{C}$
meets a single component $F_{u,0}$ of $\mathscr{X}_u$ with has $m_{u,0}=1$.
In particular in this case $m_u=1$ and all fibers are simple.

\subsubsection{Structure of simple fibers}\label{s:simple}
We focus first on the simple fibers. 
In this subsection we assume that $\pi$ has a local zero section $s$.
We start from the group theory viewpoint which
allow us to get the result quickly and produces explicit expressions.

\paragraph{The components $F_{u,i}$: group perspective.}
Let $u$ a generic point in the discriminant component $D_\alpha$.
Just as in dimension 1, the regular locus $\mathscr{X}_u^\sharp$ of the fiber $\mathscr{X}_u$
 is also an Abelian algebraic group with connected component  $F^\sharp_{u,0}$ the regular
 locus of the zero-component $F_{u,0}$.
 $\mathscr{X}_u^\sharp$ is the union of all smooth points in the components
 $F_{u,i}$ with multiplicity $m_{u,i}=1$. Hence  $\mathscr{X}_u^\sharp$ is the extension of a finite Abelian group
 by $F_{u,0}^\sharp$, and all components of multiplicity $1$ have isomorphic normalizations:
\be
m_{u,i}=1\quad\Rightarrow\quad F^\text{nor}_{u,i}\simeq F_{u,0}^\text{nor}.
\ee

By the Barsotti-Chevalley theorem \cite{MB}, $F_{u,0}^\sharp$
is a fibration over an Abelian variety, namely its Albanese variety $A_u$, with fiber a commutative group. 
Since at most two cycles vanish at the generic point $u$ (in physical language:
the degrees of freedom which get light are described by a rank-$1$ theory), 
we have $\mathsf{rank}\,H_1(A_u,\Z)=2r-2$ and 
$A_u$ is an Abelian variety of complex dimension $(r-1)$. The fiber of
$\varpi_{u,0}^\sharp\colon F^\sharp_{u,0}\to A_u$ is then a connected, non-compact, algebraic Abelian group of dimension 1, hence
a copy of either $\C$ or $\C^\times$. 
\be\label{zzxxqwa2}
1\to \C^\times \to F^\sharp_{u,0}\to A_u\to 0\qquad\text{or}\qquad
0\to \C\to F^\sharp_{u,0}\to A_u\to 0.
\ee
We call a simple singular fiber of \emph{multiplicative} (resp.\! \emph{additive}) type
if the first (resp.\! second) extension sequence applies. The group extensions of the Abelian variety $A_u$ by $\C^\times$ (multiplicative)
or respectively  by $\C$ (additive)
are parametrized by the groups \cite{serre0}
\be
\mathrm{Ext}^1(A_u,\C^\times)\simeq\mathrm{Pic}(A_u)^0\equiv A_u^\vee\quad
\text{resp.}\quad\mathrm{Ext}^1(A_u,\C)\simeq H^1(A_u,\co).
\ee

In the multiplicative case, 
$F^\sharp_{u,0}$ is the group extension defined by a point $w_u\in A_u^\vee$. We write $\cl_u\to A_u$ for the corresponding line bundle of
vanishing Chern class.  $F^\sharp_{u,0}$ is then the complement of the zero and infinity sections $s_0(0)$, $s_\infty(0)$ in total space of the bundle $\mathbb{P}(\cl_u\oplus \co)\to A_u$ \cite{sawon}. 
The normalization $F^\text{nor}_{u,0}$ of $F_{u,0}$
 is then the ruled variety $\mathbb{P}(\cl_u\oplus \co)$ over the base $A_u$.
 For example, when $r=2$ $A_u\equiv A_u^\vee$ is an elliptic curve of period $\tau$ and
the group extension defined by $w_u\in A_u$ is explicitly
\be\label{juttqwer}
F^\sharp_{u,0}\Big|_{\text{multiplicative}\atop w_u\in A_u}= \big(\C\times \C^\times\big)\Big/\Big\{
 (z,y)\sim (z+m+n\tau_u, e^{2\pi i n w_u}y),\ m,n\in\Z,\ w_u\in A_u\Big\}
\ee
We shall see in a moment that the HO invariant is the order of $w_u$ and hence is finite precisely when $w_u$ is a torsion point. The distinction $w_u$ torsion vs.\! non-torsion has an alternative characterization. Recall that an algebraic group $G$ is \emph{anti-affine} iff its ring of regular functions is trivial $\co(G)\simeq \C$. 
\begin{fact} In the multiplicative case $F^\sharp_{u,0}$ is an anti-affine group (in facts a semi-Abelian algebraic group)
iff the point $w_u$ is not torsion.
\end{fact}

In the additive case the group extension \eqref{zzxxqwa2} is specified by a $(0,1)$-class $\phi_u\in H^1(A_u,\co)\subset H^1(A_u,\C)$;
explicitly
\be
F^\sharp_{u,0}\Big|_{\text{additive}\atop \phi_u\in H^1(A_u,\co)}=(\C^{r-1}\times \C)/\Lambda_u\quad \text{where}\ \Lambda_u\ni\lambda\colon (z,y)\mapsto \big(z+\lambda,y+\langle\lambda,\phi_u\rangle\big)
\ee
where $\langle-,-\rangle\colon H_1(A_u,\Z)\times H^1(A_u,\C)\to \C$
is the natural pairing. We can see its normalization as a ruled variety over $A_u$ with a preferred $\infty$-section $s_\infty(0)$.

Translations in the Abelian variety $A_u$
are generated by holomorphic Hamiltonian vectors.\footnote{\ In classical mechanics language: the coordinates on $A_u$ are a set of angle variables for the integrable system $\mathscr{X}\to\mathscr{C}$ and the Hamiltonians $H_i$ are their canonically conjugate momenta.} The corresponding Hamiltonian functions $H_1,\dots, H_{r-1}\in\Gamma(\mathscr{C},\co)$
generate a $\C^{r-1}$ action on the fiber component $F_{u,0}$ whose kernel is the lattice $\Lambda_u$ so that $A_u=\C^{r-1}/\Lambda_u$.
This $\C^{r-1}$ action extends to the full fiber $\mathscr{X}_u$, and preserves each irreducible component $F_{u,i}$.
Then the normalization $F^\text{nor}_{u,i}$ of the $i$-th component is a
$\mathbb{P}^1$-fibration over the \emph{same} Albanese variety $A_u$ for all $i$ \cite{oguiso1}. 
The $F^\text{nor}_{u,i}$'s are ruled varieties whose lines (the fibers of the Albanese fibration $\varpi_{u,i}\colon F^\text{nor}_{u,i}\to A_u$)
are the $\C$-orbits of the Hamiltonian vector field whose Hamiltonian $h$ 
is the irreducible element in the chiral ring $\mathscr{R}\equiv\C[u_1,\dots,u_r]$ such that $D_\alpha$ has equation $h=0$
\cite{oguiso1}. Each ruled surface $ F^\text{nor}_{u,i}\to A_u$ has a finite number $k_i\geq1$ of preferred sections;
when $k_i\geq3$ we have $F^\text{nor}_{u,i}\simeq\mathbb{P}^1\times A_u$ (since the sphere with 3 punctures is rigid)
while when $k_i=2$ (resp. $1$) it may be a non-trivial fibration of the multiplicative (resp.\! additive)
kind as shown above. 
\medskip

An exceptional fiber  in codimension-1, $\mathscr{X}_u$, then has the form
\be\label{gluings}
\mathscr{X}_u=\coprod_i F_{u,i}^\text{nor}\big/\text{(identifications)}
\ee
where $F_{u,i}^\text{nor}$ are ruled varieties over $A_u$ which are glued together
along the preferred sections (with multiplicities $m_{u,i}$).
A fundamental datum of the fiber is the combinatorics of these identifications encoded in the intersection matrix.

\paragraph{The cohomology class $\eta_u$ and its physical interpretation.}
Anticipating our conclusions, we stress that, while each irreducible component $F_{u,i}$
 is a fibration over a fixed Abelian variety $A_u$, the fiber itself $\mathscr{X}_u$
 may or may not be globally a fibration over $A_u$ depending on the details of the gluings in eq.\eqref{gluings}. 
 The point is that the gluing of a section $s_i\simeq A_u$ in the $i$-th component with a
 section $s_j\simeq A_u$ in the $j$-th component is through a complex automorphism of $A_u$
 which may not  preserve the zero element, that is, be a translation\footnote{\ Non-trivial automorphisms which fix zero are ruled out for the following reason: the Lie group of $A_u$ is intrinsically identified with the Hamiltonian vector fields of the $H_i$'s and hence the identifications should act on the sections' Lie groups as the identity.}
 \be\label{kiiu8887}
 \boldsymbol{z}_j=\boldsymbol{z}_i+ \eta_{ij}\in \C^{r-1}/\Lambda_u\simeq A_u,
 \ee
 where $\boldsymbol{z}_i$ are coordinates in the base $\simeq A_u$ of $F_{u,i}$.  
 Let $\Upsilon$ be the graph dual to the components' arrangement: the $i$-th vertex of $\Upsilon$ represents
 an irreducible component $F_{u,i}$ and an edge of $\Upsilon$ between nodes $i$, $j$ stands for a section along which the corresponding components are glued. $\eta_u\equiv\{\eta_{ij}\}$ is a $1$-cocyle on $\Upsilon$ with coefficients in $\C^{r-1}/\Lambda_u$.
 Clearly we are free to redefine the coordinates $\boldsymbol{z}_i$ in each component by a translation. 
 Thus only the class $[\eta]\in H^1(\Upsilon, \C^{r-1}/\Lambda_u)$ is relevant. Clearly
 
 \begin{fact} The fiber $\mathscr{X}_u$ is a global fibration $\Pi\colon \mathscr{X}_u\to A_u$ over the
 dimension $(r-1)$ Abelian variety $A_u$ if and only if the class $[\eta]\in H^1(\Upsilon,\C^{r-1}/\Lambda_u)$ of the  gluing cocycle $\eta$ is trivial. For simple fibers \emph{HO invariant} is the order $m$ of $[\eta]$ in $H^1(\Upsilon,\C^{r-1}/\Lambda_u)$. 
 \end{fact}
\noindent In particular when $\Upsilon$ is simply connected $\mathscr{X}_u$ is a global fibration.
We shall see in a moment that when $\mathscr{X}_u\to A_u$ is a global fibration the fiber is
one of the Kodaira simple fibers.

\paragraph{Fiber dychotomy.} \emph{A priori} we may have 3 different situations:
\begin{description}
\item[(a)] the singular fiber $\mathscr{X}_u$ is not a global fibration over the Albanese variety $A_u$;
\item[(b)] the singular fiber $\mathscr{X}_u$ is a \emph{non-trivial} global  fibration over $A_u$;
\item[(c)] the singular fiber $\mathscr{X}_u$ is a product $F\times A_u$ with $F$ a Kodaira fiber and $A_u$ Abelian.
\end{description}
At first sight one would think that {\bf (b)} is the ``typical situation''; however
we shall see that \textbf{(b)} is just \emph{not possible:} the fiber is either not a (global) fibration or a trivial fibration.
The possibility \textbf{(c)} occurs precisely when the HO invariant is trivial, otherwise we have {\bf (a)}. We shall refer to this property as 
\emph{dichotomy}.
Preliminarily, we present some simple examples.

\begin{exe/c}
The simplest instance is an \emph{irreducible multiplicative fiber} (the higher dimensional analogue of Kodaira type $\mathbf{I}_1$). In this case we identify the zero and infinity sections $s_0\sim s_\infty$ of the irreducible multiplicative fiber.
E.g.\! when $r=2$ the fiber is the quotient of
 $\C\times\mathbb{P}^1$ under the identifications (for $z\in \C$, $(y_1:y_2)\in\mathbb{P}^1$)
\be
\begin{aligned}
&(z\mid y_1:y_2)\sim (z+m+n\tau_u\mid e^{2\pi i n w_u} y_1:y_2)\qquad
\forall\,m,n\in\Z\\
& (z\mid 0:y_2)\sim (z+\eta_u\mid y_1:0)
\end{aligned}
\ee
for a pair of points $w_u$, $\eta_u$ in the elliptic curve $A_u$. 
However the two points $w_u$ and $\eta_u$ are not independent: they are related by the constraints that the fiber is Lagrangian
and the polarization principal. We \textbf{Claim} that under these conditions $\eta_u\equiv w_u$.
Therefore in the irreducible multiplicative case
if the fiber is a global fibration over $A_u$, it is automatically a trivial fibration (trivial HO invariant).
\end{exe/c}

\begin{exe/c} Next example in order of complexity is a simple \emph{irreducible additive fiber} (the higher dimensional analogue of Kodaira type $\mathbf{II}$). In this case $\Upsilon$ is acyclic, $\eta_u\equiv0$,
and we remain with just the point $w_u$. We \textbf{Claim}
that the principal Lagrangian fibration condition implies $w_u=0$ i.e.\! the fiber is a product of a type $\mathbf{II}$ Kodaira fiber with $A_u$.  
\end{exe/c}

More generally 
\begin{claim} Suppose that the local monodromy $\varrho_\alpha$ around the divisor $D_\alpha$ is 
of the unstable type, $\mathrm{rank}(\varrho_\alpha-1)=2$, and the fiber over the generic point $u\in D_\alpha$
is simple; then the fiber is a product of an unstable Kodaira fiber (types $\mathbf{I}_n^*$,
$\mathbf{II}$, $\mathbf{II}^*$, $\mathbf{III}$, $\mathbf{III}^*$, $\mathbf{IV}$, $\mathbf{IV}^*$)
  and an Abelian variety $A_u$ of dimension $(r-1)$. 
\end{claim}

\paragraph{Physical intuition beyond {dichotomy.}} What is the physical distinction between situations {\bf(a)},
{\bf(b)}, and {\bf(c)}? 
Case {\bf(c)} is obvious: the quantum
system factorizes asymptotically as we approach the generic point in the discriminant  in a rank-$(r-1)$ system described by the
special geometry $\ca\to D_\alpha$ with fiber $A_u$ and an IR effective
rank-1 quantum system whose Seiberg-Witten geometry is the germ\footnote{\ The rank-1 system may be non UV complete
(this happens in the non-semisimple case); in this case the special geometry exists only in the germ sense.} of the elliptic fibration $\ce\to \Delta$
with central Kodaira fiber $F$. Situation {\bf(a)} is also clear:
although the special geometry $\ca\to D_\alpha$ and the rank-$(r-1)$ system are still well-defined \emph{per se},
the two systems do not fully decouple (asymptotically) because they remain entangled through the flux quantization  as described in \S.\,\ref{s:heuristics}.  To interprete physically situation \textbf{(b)} would be quite challenging: we have no physical observable to define and/or detect this would-be third ``quantum phase''. Physical intuition says {``just forget it, it cannot happen''}  (under our assumptions including principal polarization). This is precisely the statement of \emph{dichotomy}.

 \subsubsection{Structure of simple fibers: intersection viewpoint}
To generalize the Kodaira story from the intersection-theory side
 we may try to do
 two different things:
 \begin{itemize}
 \item[1.] define the local intersection numbers $F_{u,i}\# F_{u,j}$
 of the irreducible components along a section
  (their multiplicities $m_{u,i}$ being as in \textbf{Definition \ref{poq12zmm}});
 \item[2.] construct a maximal connected cycle $\sum_a m_a C_{i(a)}(z_a)$
of $\mathbb{P}^1$ fibers $C_{i(a)}(z_a)\subset F_{u,i(a)}$ over points $z_a\in A_u$, give appropriate definitions of their multiplicities $m_a$ and intersection numbers $C_{i(a)}\!\cdot\! C_{j(b)}$. 
\end{itemize}

As already stated, in rank-$1$ the two procedures coincide.
 
 \begin{defn} The intersection $F_{u,i}\cap F_{u,j}\subset F_{u,i}$ defines a divisor $D_j$ in $F_{u,i}$
 whose support is a section of $\varpi_{u,i}\colon F_{u,i}\to A_u$. We set
\be
F_{u,i}\# F_{u,j}\overset{\rm def}{=} C_i(z_i)\cdot D_j = C_j(z_j)\cdot D_i
\ee
where $C_i(z_i)$ is the fiber of the ruled variety $F_{u,i}^\text{nor}\to A_u$  over $z_i\in A_u$ ($F_{u,i}\# F_{u,j}$ is independent of the chosen point $z_i$ since $\C^{r-1}$ acts transitively on $A_u$). By convention
we also set $F_{u,i}\# F_{u,i}=-2$ when the fiber is reducible and 0 otherwise.
\end{defn}

\paragraph{Characteristic cycle.}  We consider two 
 fibers $C_i(z_i)$ and $C_j(z_j)$ of 
 $\varpi_{u,i}\colon F_{u,i}\to A_u$ and (respectively) $\varpi_{u,j}\colon F_{u,j}\to A_u$,
 which cross each other at a point
in the divisor 
\be
D_i\subset F_{u,i}\cap F_{u,j}\simeq A_u.
\ee 
By eq.\eqref{kiiu8887} in order to cross they must be of the form
$C_i(z_i)$ and $C_j(z_i+\eta_{ij})$.
We define the (local) intersection number
of the two rational curves, 
\be
C_i(z_i)\# C_j(z_i+\eta_{ij})\overset{\rm def}{=}F_{u,i}\# F_{u,j}.
\ee
The multiplicity $m_{i,z_i}$ of the rational curve
$C_i(z_i)$ is defined to be the multiplicity
$m_{u,i}$ of the component $F_{u,i}$ which contains it.
A maximal connected cycle of $\mathbb{P}^1$ fibers 
\be\label{vu}
\boldsymbol{\Psi}=\sum_{v\in V_u} m_{u,i(v)} C_{i(v)}(z_v)
\ee
 is
called the \emph{characteristic cycle} of the exceptional fiber \cite{oguiso1,oguiso2,oguiso3}.
We stress that the two index sets in eqs.\eqref{iu}, \eqref{vu}, $I_u$ and $V_u$, need \emph{not} to be equal.

The arguments used by Kodaira for the elliptic fibers apply both to the
fiber cycle $\mathscr{X}_u=\sum_{i\in I_u} m_{u,i} F_{u,i}$ and to the characteristic cycle
$\boldsymbol{\Psi}$. In particular, both intersection matrices 
\be
C_{i,j}\equiv F_{u,i}\# F_{u,j},\qquad K_{v,w}\equiv C_{i(v)}(z_v)\# C_{i(w)}(z_w),
\ee
 should be either zero or the negative of the Cartan matrix of a simple, simply-laced affine
 Lie algebra of type $\widetilde{A}_{n-1}$, $\widetilde{D}_{4+n}$,
$\widetilde{E}_6$, $\widetilde{E}_7$ or $\widetilde{E}_8$, while
for the Cartan matrices of $\widetilde{A}_1$ and $\widetilde{A}_2$
we have two  different realizations akin to rank-1 case: for $\widetilde{A}_1$
the two components may either cross transversely at two distinct sections,
or tangentially at a single section; for $\widetilde{A}_2$ the three components
may meet pairwise transversely at three distinct points or all at the same point.
The conclusion is that both the fiber cycle $\sum_i m_{u,i} F_{u,i}$
and the characteristic cycle $\boldsymbol{\Psi}$ are ``modelled'' on a Kodaira fiber,
\emph{except that} $\boldsymbol{\Psi}$, contrary to the fiber cycle,
may have \emph{infinitely many} components so that the type
$\mathbf{I}_\infty$ is also allowed for the characteristic cycle.\footnote{\ A more careful analysis \cite{oguiso2}
shows that Type $\mathbf{I}^*_\infty$ is ruled out.}
Cycles modelled on a Kodaira fiber have (by definition)
 the same number of components, intersection form, and multiplicities as their model Kodaira fiber.

\begin{rem} The characteristic cycle type is well-defined, i.e.\! independent of the various choices.
Indeed, $\C^{r-1}$ acts transitively on the set of maximal connected cycles of $\mathbb{P}^1$ fibers.\end{rem}

\begin{rem} One extends the allowed Kodaira type of the characteristic cycle to type $\mathbf{I}_0$
(relevant only when the fiber is multiple). In this case $\boldsymbol{\Psi}$ is a smooth elliptic curve.
\end{rem}

However, in general, the Kodaira type of  the characteristic cycle
$\boldsymbol{\Psi}$ is \emph{different} from the Kodaira type of the fiber cycle
$\mathscr{X}_u$.  

\begin{claim} The Kodaira types of a \emph{simple} fiber and of its characteristic cycles
coincide if and only if the class $[\eta]\in H^1(\Upsilon,A_u)$ is trivial.
\end{claim}

In particular when the simple fiber is reducible, and the fiber cycle is not one
of the afore mentioned special types, $\Upsilon$ is the dual of the 
affine Dynkin graph associated to the fiber cycle, which is acyclic, 
except for $\widetilde{A}_{b-1}$ i.e.\! type $\mathbf{I}_b$. 
Therefore for simple fibers $[\eta]$ may be non-zero only if the fiber cycle has type $\mathbf{I}_b$
with $b\geq1$, i.e.\! for \emph{stable} singular fibers.
 It is easy to check that the statement extends to the special cases.

\subsubsection{Simple stable fibers: explicit expressions}\label{s:explicit}
Recall that the polarization of the special geometry is assumed
to be principal. In a sufficiently small neighborhood of a generic point $u$ in a \emph{stable} discriminant
$D_\alpha$ we can find special holomorphic coordinates $a^i$ ($i=1,\cdots,r$)
with local discriminant $a^r=0$ and local prepotential of the form
\be\label{poioqw12}
\cf(a^i)= \widetilde{\cf}(a^i)+\frac{\ell}{4\pi i}(a^r)^2\log a^r,\qquad \ell\in\mathbb{N}.
\ee
While this is physically obvious (it corresponds to some hypers with electric charges $q_i$ getting massless, while $\ell=\sum_i q_i^2$), the mathematically oriented reader may wish to have a look to \cite{oguiso3} for a math proof.
We write
\be
\tau_{ij}\overset{\rm def}{=}\frac{\partial^2\widetilde{\cf}}{\partial a^i\partial a^j}\bigg|_u,\qquad \theta_i\overset{\rm def}{=} \frac{\partial^2\widetilde{\cf}}{\partial a^i\partial a^r}\bigg|_{u}\quad i,j=1,\dots,r-1.
\ee
$\tau_{ij}$ is the period matrix of the Abelian variety $A_u$ in the chosen duality frame (marking in the math jargon). In other words, writing $\Lambda_u$ for the lattice 
\be
\Lambda_u\equiv \Big\{m_i+\tau_{ij}\mspace{1mu} n^j\colon m_i,n^j\in\Z^{r-1}\Big\}\subset\C^{r-1},
\ee
 we have $A_u=\C^{r-1}/\Lambda_u$. In the simple stable case all components
 are multiplicative of multiplicity 1, and as an algebraic group\footnote{\ Here we use that the integer $\ell$ in eqn.\eqref{poioqw12} is the number of components in the fiber just as in Kodaira's rank-1 case. This can be inferred by starting from the particular case $\theta=0$ where the statement is obvious (since the fiber is the product of $A_u$ and a Kodaira fiber of type $\mathbf{I}_\ell$) and then deform away from that point making $\theta$ non-zero. Clearly the deformation would not change the monodromy or the number of components. Readers not satisfied with this argument are referred to \cite{oguiso3}.}
 \be
 F_u^\sharp = F_{u,0}^\sharp \rtimes \Z_\ell.
 \ee 
 In particular the normalized components $F_{u,i}$ are all isomorphic copies of a ruled variety $\mathbb{P}(\cl\oplus\co)\to\C^{r-1}/\Lambda_u$.
 We write $z_i\in \C^{r-1}/\Lambda_u$ for the
coordinate in the $A_u$ basis of the $i$-th copy.
 We think $\theta\equiv (\theta_i)\bmod\Lambda_u$ as a point in $A_u$.
 \medskip
 
Let $\boldsymbol{Z}$ be the \emph{infinite} cycle of copies $\{\varpi_{u,i}\colon F_{u,i}\to A_u\}_{i\in\Z}$ of the
 ruled variety with preferred zero- and $\infty$-sections,
 $s_0(i),s_\infty(i)\colon A_u\to F_{u,i}$ obtained by gluing the $0$-section of $F_{u,i+1}$
 to the $\infty$-section of $F_{u,i}$. 
  \be
 \boldsymbol{Z}=\coprod_{i\in \Z} F_{u,i}\Big/\big(s_0(i+1)\sim s_\infty(i)\big).
 \ee
 Since the dual graph of $\boldsymbol{Z}$, $A^\infty_\infty$, is acyclic
 we may assume with no loss that the gluing is via the identity map $A_u\to A_u$, i.e.\! $z_{i+1}=z_i$.
  On $\boldsymbol{Z}$ we have a maximal connected cycle of countably-many rational curves
  \be\label{kiuuqert}
  \boldsymbol{\widetilde{\Psi}}= \sum_{i\in\Z} C_i(0),
  \ee 
  and all other maximal cycles are obtained by acting with $\C^{r-1}$ on it.

 The fiber $\mathscr{X}_u$ with pre-potential \eqref{poioqw12} is the quotient of $\boldsymbol{Z}$ 
 under the identification \cite{oguiso3}
 \be
 (F_{u,i}, z_i) \simeq (F_{u,i+\ell}, z_i+\theta)
 \ee
 In other words: we have $\ell$ distinct components $F_{u,i}$ cyclically identified as in figure 2
 of \cite{sawon} (see also figure 1 of \cite{jap1}),
 and the fiber cycle has Kodaira type $\mathbf{I}_\ell$.
 However when we  go around
 the components' cycle following a fiber (i.e.\! an orbit of the Hamiltonian flow generated by $a^r$)  we do not come back where we started from, but rather to a point of $s_0(0)\simeq A_u$ translated by $\theta$.
 In particular, the components of the maximal 1-cycle \eqref{kiuuqert} do not get identified mod $\ell$,
 unless $\theta=0\bmod\Lambda_u$. Indeed 
 \be
 C_{i+\ell}(0)=C_i(\theta)\ \big(\neq C_i(0)\ \text{for }\theta\neq0\big).
 \ee
 If $\theta$ is a $m$-torsion point, i.e.\! $m$ is the minimal positive integer such that
 $m\mspace{1mu}\theta=0\bmod\Lambda_u$, the rational curve will return to itself after
 going around $m$-times the components' cycle
 \be
 C_{i+m\ell}= C_i(m\theta)\equiv C_i(0).
 \ee
  In this case the characteristic cycle is
 \be
 \boldsymbol{\Psi}=\sum_{s=0}^{m-1}\sum_{i=0}^{\ell-1} C_{i}(\ell\theta)
 \ee
 with $m\ell$ components and Kodaira type $\mathbf{I}_{m\ell}$. If $\theta\in A_u$ is not a torsion point, the
 characteristic cycle will wrap around the fiber cycle indefinitely, and the Kodaira type of the
 characteristic cycle will be $\mathbf{I}_\infty$. For a visualization of the fiber geometry we refer the reader to the 
 beautiful figures in ref.\cite{sawon}.

 \paragraph{The line bundle $\cl\to A_u$.} It remains to show our claim that in the irreducible multiplicative case $w_u\equiv \eta_u$, or for the more general stable case that the line bundle $\cl$ such that $F_{u,i}\simeq\mathbb{P}(\cl\oplus\co)$ coincides with the point $\theta\in A_u^\vee\simeq A_u$. 
 Notice that for $a^r$ small but non-zero the fiber is a smooth Abelian variety of dimension $r$
 with period matrix $\partial^2\cf/\partial a^I\partial a^J$. If $(z_i,y)$ are coordinates in the covering
 $\C^r$ we have (in particular!) the identifications ($i,j=1,\dots, r-1$)
 \be
 (z_i,y)\sim (z_i+m_i +\tau_{ij} n^j, y+m+\theta_i n^i)\quad m_i,m,n^i\in\Z
 \ee 
 or, in the multiplicative notation,
 \be
 (z_i,w)\equiv (z_i, e^{2\pi i y}) \sim (z_i+m_i+\tau_{ij} n^j, e^{2\pi i \theta_i n_i}w),
 \ee
which is (the generalization of) eq.\eqref{juttqwer}.

\paragraph{The HO invariant.} In the simple stable case we define the HO invariant to be the number of times the characteristic cycle wraps around the fiber cycle.  In the case of simple unstable fibers the graph $\Upsilon$ is acyclic,
 and the Kodaira types of the characteristic cycle and fiber cycle
 are equal and belong to the list
 \be
 \mathbf{I}^*_b,\ \mathbf{II},\ \mathbf{III},\ \mathbf{IV},\ \mathbf{II}^*,\ \mathbf{III}^*,\ \mathbf{IV}^*.
 \ee
 
We define the HO invariant for \emph{simple fibers}
to be the ratio
\be\label{asdefined}
\mu(u)=\frac{\text{(number of components of the fundamental cycle)}}{\text{(number of irreducible components of the fiber)}}\in\mathbb{N}\cup \infty
\ee 
The HO invariant is trivial precisely when the Kodaira type of the characteristic cycle agrees with the one of the fiber cycle. In this case the singular fiber is a product $K\times A_u$ with $K$ the Kodaira fiber of their common type and
$A_u$ a polarized Abelian variety of dimension $(r-1)$.

\subsubsection{Kodaira type vs. monodromy} As in rank-1, we can describe the situation in terms
of the local monodromy $\varrho_\alpha$ around $D_\alpha$. The $2r-2$ homology classes corresponding to $H_1(A_u,\Z)$
are invariant under the local monodromy, which then lays in a $SL(2,\Z)$ subgroup of $Sp(2r,\Z)$,
and can be identified with a Kodaira monodromy (tables \ref{table}, \ref{table2}). In particular
\be
(\varrho_\alpha^m-1)^2=0\quad\text{where }m=1,2,3,4,6\quad\text{and}\quad m=3,4,6\ \Rightarrow\ \varrho_\alpha^m=1.
\ee

\begin{claim}\label{poiqwerc} Suppose the fiber over the generic point $u\in D_\alpha$ is \textbf{simple}.
The conjugacy class of the local monodromy $\varrho_\alpha$ around $D_\alpha$ 
 is given by the Kodaira
monodromy associated to the Kodaira type of the \textbf{fiber cycle} of the fiber $\mathscr{X}_u$
at a generic point $u\in D_\alpha$.
\end{claim}
\noindent We stress that it is the fiber cycle and not the characteristic cycle which controls the local monodromy.
Indeed the local monodromy is independent (modulo conjugacy) of the point $u$, whereas the characteristic
cycle may depend on $u$ in a very wild way, cf. \S.\,\ref{s:wild}.

We show \textbf{Claim \ref{poiqwerc}}.
Let $\Delta=\{(h_1,\dots,h_r)\colon |h_i|<\epsilon\}\subset \mathscr{C}$ be a polydisk
centered at $u\in D_\alpha$ with $h_r=0$ the local equation of $D_\alpha$ in $\Delta$
and $\mathscr{X}|\to \Delta$ the restricted Lagrangian fibration.
Pulling back the fibration through the branched cover 
\be\label{covtrick}
f\colon\widetilde{\Delta}\to\Delta, \qquad f\colon h_r\mapsto h_r^m
\ee 
we get a new fibration $\widetilde{\mathscr{X}}|\to \widetilde{\Delta}$ whose fibers over $\widetilde{\Delta}^\sharp\equiv\widetilde{\Delta}\setminus(\widetilde{h}_r)$ ($\widetilde{h}_r\equiv f^*h_r$)
are smooth principally polarized Abelian varieties. The family of smooth
Abelian varieties over the complement of the discriminant
 \be\label{fffafafq}
 \widetilde{\mathscr{X}}|_{\widetilde{\Delta}\setminus (\widetilde{h}_r)}\to \widetilde{\Delta}\setminus (\widetilde{h}_r),
 \ee
has now a strictly unipotent monodromy $\widetilde{\varrho}_\alpha$
around the discriminant $(\widetilde{h}_\alpha)$: 
\be
(\widetilde{\varrho}_\alpha-1)^2=0.
\ee 
In other words (as it is well known \cite{vhs4,vhs5,vhs6,periods}) via the covering trick \eqref{covtrick}
 we can always
 reduce ourselves to the stable situation: see \cite{oguiso2} for the application of the trick in the present context.

Let us consider first the case where original monodromy $\varrho_\alpha$ was \emph{semisimple}, i.e.\!
$\varrho_\alpha^m=1$ with $m\in\{2,3,4,6\}$.
The pulled back monodromy is now trivial $\widetilde{\varrho}_\alpha=1$, and we may extend the family of
Abelian varieties \eqref{fffafafq} along the discriminant $h_r=0$ to get a fibration $\widetilde{\mathscr{X}}\to\widetilde{\Delta}$ (see e.g. \textbf{Theorem $\boldsymbol{13\!\cdot\!7\!\cdot\!5}$} of \cite{periods}). 
The central fiber
$\widetilde{\mathscr{X}}_u$ of the family is now a principally polarized, $r$-dimensional, Abelian variety with a polarized
automorphism $\xi$ of order $m$ whose analytic representation\footnote{\ Let $\xi\colon A\to A$ an automorphism of the Abelian variety of dimension $r$ (over $\C$). Its \emph{analytic representation}
\cite{complexabelian} is the induced complex representation on the tangent space
at the origin $T_0A\simeq \C^r$ which is the Lie algebra $\mathsf{Lie}(A)$ of the complex Lie group $A$.} $\rho_a(\xi)$ has $(r-1)$ eigenvalues $1$ while its last eigenvalue is a primitive $m$-th root of unity.
By (say) \textbf{Theorem 13.2.8} of \cite{complexabelian}, there exist two polarized Abelian subvarieties of $\widetilde{\mathscr{X}}_u$, $A_u^\vee$ and $E_u$,
respectively of dimension $(r-1)$ and $1$ and an isogeny
\be\label{juuyqw}
A_u^\vee \times E_u\to \widetilde{\mathscr{X}}_u, \qquad\text{with}\quad \xi\in\mathsf{Aut}(E_u).
\ee  
 Since $\widetilde{\mathscr{X}}_u$ is principally polarized, we may identify $A_u^\vee$ with the dual of the Albanese
 variety $A_u$ i.e.\! the Picard variety of $\mathscr{X}_u$.  $A_u^\vee\not\simeq A_u$ in general, i.e.\! $A_u$ needs not to be \emph{principally} polarized. In particular, when $m=3,\;4,\;6$ the complex structure of the elliptic curve $E_u$ is fixed 
 to $\tau= e^{2\pi i/3}$, $i$, and $e^{2\pi i/3}$ respectively. When $m=2$, $E_u$ can be any elliptic curve.
When $m\in\{3,4,6\}$ the field $\mathbb{Q}(e^{2\pi i/m})$ has degree $2$, and 
we have two choices for the CM-type of $E_u$  \cite{complexabelian}:
as discussed in \cite{caorsi} they correspond to the two possible Kodaira types with a semisimple monodromy of the given order $m$ namely
\begin{itemize}
\item
 $\mathbf{IV}$, $\mathbf{IV}^\ast$ for $m=3$;
 \item $\mathbf{III}$, $\mathbf{III}^\ast$ for $m=4$;
 \item $\mathbf{II}$, $\mathbf{II}^\ast$ for $m=6$.
 \end{itemize}
 
 Taking the quotient of $\widetilde{\mathscr{X}}\to\widetilde{\Delta}$ by $\Z_m$ we return to a fibration over
 the original polydisk $\Delta$,
  $\mathscr{X}^\flat\to \Delta$ whose restriction to $\Delta\setminus (h_\alpha)$ is isomorphic to the
  restriction of our original fibration   $\mathscr{X}\to \Delta$.
The central fiber of $\mathscr{X}^\flat$ has the form $m F$ where the reduced irreducible variety  
 $F$ is a ruled variety over $A_u$. However $\mathscr{X}^\flat$ is not smooth: to get a smooth
 total space we need to perform an appropriate sequence of blow-up and blow-down of various components
 inside the central fiber. The precise sequence 
  of these operations for each semisimple Kodaira type can be read in \S.\,4 of \cite{oguiso2}
 (or \S.\.V.10 of \cite{bhm}). In particular we conclude that in the semisimple case the order
 of the local monodromy agrees with the Kodaira type of the exceptional fiber which is a product
 of a Kodaira elliptic singular fiber $K$ times $A_u$.

The covering trick (with $m=2$) reduces a fiber of type $\mathbf{I}^*_n$ to a fiber of type $\mathbf{I}_{2n}$
 (again after suitable blow-downs/ups of components of the central fiber). This shows that \textbf{Claim \ref{poiqwerc}} is true
 iff it holds for the stable case. The stable case was studied in great detail in \S.\,4 of \cite{oguiso3}
 (assuming a principal polarization). There it is shown that a stable fiber with
 $\ell$ components has a local monodromy conjugate over $Sp(2r,\Z)$ to
 \be
 \left(\begin{array}{c|c}\;\begin{smallmatrix}1 & \ell\\ 0 &1\end{smallmatrix}\;  & 0\\\hline
 0 & \boldsymbol{1}_{r-1}\end{array}\right).
 \ee
 This completes the argument.

 \subsection{Multiple fibers}\label{s:multiple}
 
Next we consider \emph{multiple} Lagrangian fibers following \cite{oguiso2}. 
Multiple fibers may be present only in a ``no-section'' special geometry.
As discussed in  \S.\,\ref{s:preliminary} the ``no-section'' geometries have not the physical interpretation of standard Seiberg-Witten geometries: their physical meaning will be discussed in section \ref{s:general}.
\medskip

In a Lagrangian fibration of a holomorphic symplectic manifold only a handful of
 multiplicities $m\geq2$ are allowed, and only for specific types of the characteristic cycle $\boldsymbol{\Psi}$.
\textbf{Theorem 1.1} of \cite{oguiso2} gives the following table of allowed multiple fibers:
\be\label{multiple}
\begin{tabular}{c|l@{\quad\ }cccc}\hline\hline
multiplicity $m$ & \multicolumn{5}{c}{characteristic cycle type}\\\hline
$6$ & $\mathbf{I}_0$ ($E_{\zeta_3}$)\\
$5$ & $\mathbf{II}$\\
$4$ & $\mathbf{I}_0$ ($E_{\zeta_4}$) & $\mathbf{IV}$\\
$3$ & $\mathbf{I}_0$ ($E_{\zeta_3}$) & $\mathbf{III}$ & $\mathbf{I}_0^*$\\
$2$ & $\mathbf{I}_{2n}$ ($n\geq0$) & $\mathbf{I}_\infty$ & $\mathbf{I}_0^*$ & $\mathbf{IV}$ & $\mathbf{IV}^*$\\\hline\hline
\end{tabular}
\ee
The notation $\mathbf{I}_0$ $(E_{\zeta_n})$ means that the $\mathbf{I}_0$ type characteristic cycle is required to be an elliptic curve whose period $\tau$
is a primitive $n$-th root $\zeta_n$ of 1.\footnote{\ Throughout this paper $\zeta_n$ will always denote
a primitive $n$-th root of unity.}  

\paragraph{Local monodromy around a multiple fiber.} As it will be clear from the explicit construction below, in the semisimple case
the local monodromy $\varrho_\alpha$ satisfies (for $m\geq2$)
the following equation
\be
[\varrho_\alpha^m]=\begin{smallmatrix}\textbf{conjugacy class of the local monodromy}\\ \textbf{ around the Kodaira fiber on which the}\\
\textbf{characteristic cycle is modelled}\end{smallmatrix}
\ee
that is, the local monodromy is the one of the Kodaira fiber associated to the Lie algebra $\mathfrak{g}$,
 see the second column of table \ref{equuua}. In the non-semisimple case $m=2$
 $\boldsymbol{\Psi}=\mathbf{I}_{2n}$ we have $\mathfrak{g}=D_{4+\ell}$
 where $2\ell\mid 2n$ is the number of components of the fiber.
 We notice that monodromies of types $\mathbf{II}$, $\mathbf{III}$
and $\mathbf{IV}$ (whose dual graph is not an affine Dynkin diagram) cannot appear.

\begin{table}
\begin{center}
\begin{tabular}{c|c|c|c}\hline\hline
$(m,\boldsymbol{\Psi}\ \text{type})$ & monodromy & equation of $\ce$ & $\Z_m$ automorphism $\sigma$\\\hline
$(6,E_{\zeta_3})$ & $E_8$ & $y^2=x^3+1$ & $(x,y,t)\mapsto (\zeta_6^2 x,\zeta_6^3 y,\zeta_6 t)$\\
$(5,\mathbf{II})$ & $E_8$ & $y^2=x^3+t$ &  $(x,y,t)\mapsto (\zeta_5^2 x,\zeta_5^3 y,\zeta_5 t)$\\
$(4,E_{\zeta_4})$ & $E_7$ & $y^2=x^3+x$ & $(x,y,t)\mapsto (- x,\zeta_4^3 y,\zeta_4 t)$\\
$(4,\mathbf{IV})$ & $E_8$ & $y^2=x^3+t^2$ & $(x,y,t)\mapsto (- x,\zeta_4^3 y,\zeta_4 t)$\\
$(3,E_{\zeta_3})$ & $E_6$ & $y^2=x^3+1$ & $(x,y,t)\mapsto (\zeta_3^2 x, y,\zeta_3 t)$\\
$(3,\mathbf{III})$ & $E_7$ & $y^2=x^3+t x$ & $(x,y,t)\mapsto (\zeta_3^2 x, y,\zeta_3 t)$\\
$(3,\mathbf{I}_0^*)$ & $E_8$ & $y^2=x^3+t^3$ & $(x,y,t)\mapsto (\zeta_3^2 x, y,\zeta_3 t)$\\
$(2,\mathbf{I}_0)$ & $D_4$ & $y^2=x^3+ax+b$ & $(x,y,t)\mapsto ( x, -y,-t)$\\
$(2,\mathbf{I}_0^*)$ & $E_7$ & $y^2=x^3+t^2 x$ & $(x,y,t)\mapsto ( x, -y,-t)$\\
$(2,\mathbf{IV})$ & $E_6$ & $y^2=x^3+t^2$ & $(x,y,t)\mapsto ( x, -y,-t)$\\
$(2,\mathbf{IV}^*)$ & $E_8$ & $y^2=x^3+t^4$ & $(x,y,t)\mapsto ( x, -y,-t)$\\\hline\hline
\end{tabular} 
\caption{\label{equuua} Data for the construction of multiple Lagrangian fibers. The $2^\text{nd}$ column specifies the local monodromy in terms of the corresponding Lie algebras $\mathfrak{g}$ as listed in  tables \ref{table}, \ref{table2}.}
\end{center}
\end{table}

\subsubsection{Explicit construction}\label{s:explicit} 
Ref.\cite{oguiso2} gives an explicit construction in rank 2
of all allowed multiple
fibers in table \eqref{multiple}. For sake of comparison with physics we review their construction. We first 
assume that the characteristic cycle is not of type $\mathbf{I}_{2n}$ with $n>0$.

For each allowed pair $(m,\boldsymbol{\Psi}\ \text{type})$ in the first column of table \ref{equuua}, we write $\ce\to \C$ for the elliptic fibration whose central fiber $\ce_0$ has the same Kodaira type as $\boldsymbol{\Psi}$: see third column of 
table \ref{equuua} for the explicit Weierstrass model of each fibration ($t$ is the coordinate on the base $\C$).
The rank-1 special geometry $\ce$ has a natural symplectic form
\be
\omega=\frac{dx\wedge dt}{y}.
\ee
The fibration $\ce\to \C$ has a $\Z_m$ automorphism\footnote{\ For the interpretation of this automorphism
in the 4d/2d correspondence, and its relation to the quantum monodromy of the 4d $\cn=2$ QFT, see \cite{Cecotti:2010fi,Cecotti:2014zga}.} $\sigma$
acting on the base coordinate as $t\mapsto \zeta_m\, t$ and preserving $\omega$. 
Let $E$ be a \emph{fixed} elliptic curve (with holomorphic differential $dz$) and $p_m\in E$
an $m$-torsion point. Consider the fibration
\be
\mathscr{Y}\equiv\ce \times E\times \C \to \C\times \C,\quad \big((x,y,t),z,s\big)\mapsto (t,s),
\ee
with  symplectic form
\be
\Omega=\omega+dz\wedge ds.
\ee
Take the quotient of $\mathscr{Y}$ with respect to the $\Z_m$ group generated by the order-$m$ symplectic automorphism
\be\label{seeeg}
\Sigma\colon \big((x,y,t),z,s\big)\mapsto \big((\zeta_m^2 x, \zeta_m^3 y, \zeta_m t),z+p_m,s\big).
\ee
The resulting Lagrangian fibration
\be
\mathscr{X}\equiv \mathscr{Y}/\Z_m\longrightarrow \C/\Z_m\times \C\simeq \C^2
\ee
has a central fiber of multiplicity $m$ and type $\boldsymbol{\Psi}$ \cite{oguiso2}.
The Albanese variety of an irreducible component of the singular fiber is $E/\Z_m$.
The above construction has obvious generalizations to higher dimensions.

%

\paragraph{Multiple fibers with $\mathbf{I}_{2n}$ characteristic cycle.}
In the $\mathbf{I}_{2n>0}$ case
we have the extra feature of the possibility of a non-trivial HO invariant in the sense
of our discussion in \S.\,\ref{s:explicit}. A part for that, the construction for $\mathbf{I}_{2n}$ is pretty similar to the one for the other cases;
the explicit geometry is constructed in \textbf{Example 6.3} of \cite{oguiso2}. Note that we may also have $n=\infty$.

\begin{rem} Fibers of multiplicity 2 and characteristic cycle of type $\mathbf{I}_{2n}$ ($n>0$) have an even number of components (i.e.\! their fiber cycle is of type $\mathbf{I}_{2n^\prime}$ with $n^\prime\mid n$).
\end{rem}

\begin{rem}
We stress that we have double fibers of characteristic cycle type $\mathbf{I}_{2n}$ with any HO invariant
$\mu$ (as defined in eq.\eqref{asdefined}).
\end{rem}

\subsubsection{Deformations spaces of multiple Lagrangian fibers}\label{defspaces}

``(Partially) frozen'' singularities have a deformation space $\cm$ of smaller dimension than naively expected
\be
\dim_\C\cm < \mathrm{rank}\,\mathfrak{g}
\ee
where $\mathfrak{g}$ is the Lie algebra associated to the conjugacy class of the local monodromy.
In order to show that our geometrical description is on the right track, we need to construct \emph{geometrically} the deformation space and check that it has precisely the dimension predicted by physics for the (partially) frozen singularity of each given kind. To match the physical expectation the dimension should be equal to the rank of the corresponding flavor group
as described in the following two physical sections. 

For definiteness we take $r=2$. The Hwang-Oguiso construction (\S.\,\ref{s:explicit}) of the multiple fiber started from a covering family of elliptic curves
with Weierstrass equation of the form
\be
y^2=P_0(x,t)\quad\text{where}\quad P_0(\zeta_m^2\mspace{2mu} x,\zeta_m\mspace{2mu} t)=\zeta_m^6\mspace{2mu} P_0(x,t),
\ee
which are listed in table \ref{equuua} for each $\boldsymbol{\Psi}$'s Kodaira type and multiplicity $m$. 
A deformation of the geometry is a deformation of the polynomial
\be
P_0(x,t)\leadsto P(x,t)
\ee  
subjected to the following rules:
 \begin{itemize}
 \item[\textit{(i)}] the deformation does not change the geometry at infinity, i.e.\! the deformation consists in adding to $P_0(x,t)$  monomials $x^at^b$ with $(a,b)\in\Z^2\subset\R^2$ contained in the closure of the Newton polytope $N$ of $P_0(x,t)$,
the convex hull $N\subset\R^2$ of the points $\{(0,0),(a_1,b_1),\dots,(a_s,b_s)\}$ where 
 $(a_i,b_i)$ are the degrees of the monomials in $P_0(x,t)$;
\item[\textit{(ii)}] the deformation must preserve the symmetry $\sigma$
 \be
 P(\zeta_m^2\mspace{2mu} x,\zeta_m\mspace{2mu} t)=\zeta_m^6\mspace{2mu} P(x,t).
 \ee
 \item[\it(iii)] two polynomials $P(x,t)$, $P^\prime(x,t)$ which satisfy \textit{(i)} and \textit{(ii)} are considered to be equivalent
 if they are related by a linear redefinition of the variables $P(x,t)^\prime=P(x+Q(t),t)$ for some polynomial $Q(t)$. 
 \end{itemize}
 For instance, for the pair $(2,\mathbf{IV}^*)$ the Newton polytope in $\R^2$ is
\be\label{4445rt}
\begin{gathered}
\includegraphics[width=0.15\textwidth]{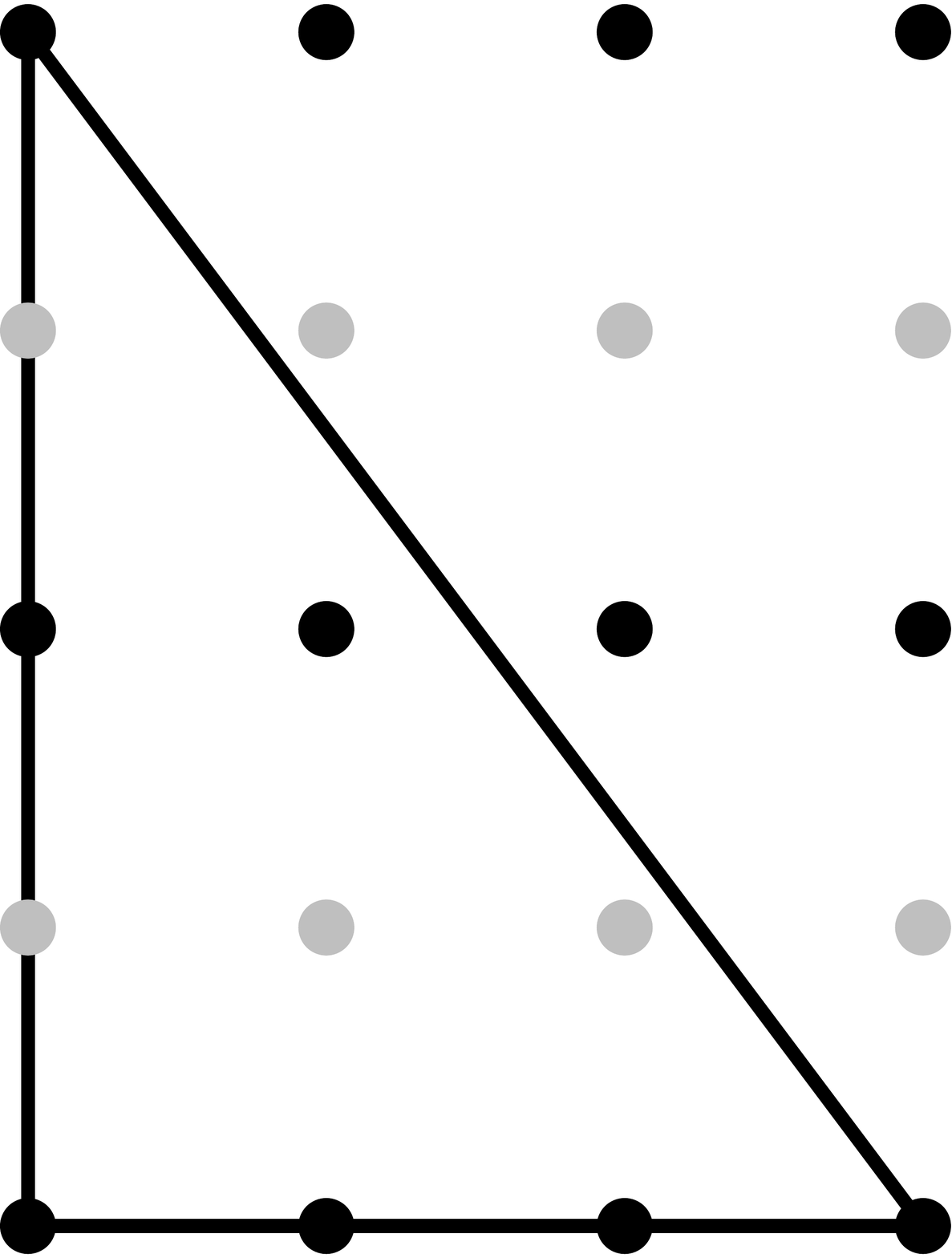}
\end{gathered}
\ee
(black/gray dots stand for $\Z^2\subset\R^2$). The monomials satisfying \textit{(i)} are $1$, $x$, $x^2$, $x^3$, $t$, $tx$, $tx^2$, $t^2$, $t^2x$, $t^3$, and $t^4$,
 while the monomials satisfying \textit{(ii)} are the ones containing $t$ to an even power (black nodes in \eqref{4445rt}). We remain with
 \be
 P(x,t)= x^3+ b_1 x^2+ b_2 x + b_3 + b_4 t^2 x+ b_5 t^2+ t^4,
 \ee
while \textit{(iii)} the redefinition $x\leadsto x-b_1/3$ sets the coefficient of $x^2$ to zero,
leaving with a deformation space $\cm$ of dimension $4$.
  In table \ref{equuua2} we list the polynomials describing 
 the deformations of each multiple fiber: the $c_i$'s are the effective deformation parameters.

\begin{table}
\begin{center}
\begin{tabular}{c|c|c|c}\hline\hline
$(m,\boldsymbol{\Psi}\ \text{type})$ & equation of $\ce$  & deformed equation & $\dim_\C\cm$\\\hline
$(6,E_{\zeta_3})$ & $y^2=x^3+1$  & $y^2=x^3+1$ & $0$\\
$(5,\mathbf{II})$ & $y^2=x^3+t$  & $y^2=x^3+t$ & $0$\\
$(4,E_{\zeta_4})$ & $y^2=x^3+x$  & $y^2=x^3+x$ & $0$\\
$(4,\mathbf{IV})$ & $y^2=x^3+t^2$ & $y^2=x^3+c_1 x+t^2$ & $1$\\
$(3,E_{\zeta_3})$ & $y^2=x^3+1$  & $y^2=x^3+1$ & $0$\\
$(3,\mathbf{III})$ & $y^2=x^3+t x$  & $y^2=x^3+t x+c_1$ & $1$\\
$(3,\mathbf{I}_0^*)$ & $y^2=x^3+t^3$ & $y^2=x^3+a_1 x t+ t^3+a_2$ & $2$\\
$(2,\mathbf{I}_0)$ & $y^2=x^3+Ax+B$  & $y^2=x^3+Ax+B$ & $0$\\
$(2,\mathbf{I}_0^*)$ & $y^2=x^3+t^2 x$  & $y^2=x^3+c_1 x^2 +c_2 x+ c_3 +  t^2 x$ & $3$\\
$(2,\mathbf{IV})$ & $y^2=x^3+t^2$  & $y^2=x^3+c_1 x +c_2+t^2$ & $2$\\
$(2,\mathbf{IV}^*)$ & $y^2=x^3+t^4$  & $y^2=x^3+(c_1+c_2t^2)x+(c_3+c_4 t^2)+t^4$ & $4$\\\hline\hline
\end{tabular} 
\caption{\label{equuua2} Deformations of the covering Weierstrass equation}
\end{center}
\end{table}

\section{Physics of ``no-section'' special geometries}\label{s:general}

We now consider the physical interpretation of special geometries which are Lagrangian fibrations
$\mathscr{X}\to\mathscr{C}$ \emph{without} a global section.
This opens up the possibility of multiple singular fibers. The allowed multiple fibers
are listed in table \eqref{multiple}
following ref.\cite{oguiso2}.

As explained in \S.\,\ref{s:preliminary} these special geometries cannot have the interpretation of standard Seiberg-Witten
geometries on ``usual sense'' Coulomb branches. Their physical interpretation is rather subtle. The easiest way to see that these more general geometries \emph{do describe} actual supersymmetric quantum system is to relate them to M-/F-theory constructions. We start from this perspective.   

\subsection{Comparison with M- and F-theory}\label{s:poiuqw}

Looking at the table \eqref{multiple} of multiple Lagrangian fibers in rank $r\geq2$
one cannot not notice an intriguing feature:
the Lagrangian multiple fibers  are in one-to-one correspondence\footnote{\ For the precise sense in which the correspondence is 1-to-1 see \textbf{Remark \ref{yyyyttt2}}.}
with the (partially) frozen singularities in M-theory (in complex codimension 2) \cite{F1,F2,F3,MM1,MM2}! 
In turn these frozen singularities are in one-to-one correspondence with the ``non-obvious'' components of the moduli spaces $\cn_G$ of
flat $G$-connections on a 3-torus $T^3$ \cite{borel,W4}, that is, components different from the connected component of the
trivial connection
\be
\cn_G^0=(\boldsymbol{T}_G\times\boldsymbol{T}_G\times \boldsymbol{T}_G)\big/\mathsf{Weyl}(G)
\ee  
(here $\boldsymbol{T}_G\subset G$ is the maximal torus and $\mathsf{Lie}(G)=\mathfrak{g}$). For a given $G$
we have \cite{borel}
\be
\cn_G=\mspace{-20mu}\coprod_{m\in \mathbf{C}(\mathfrak{g})\atop n\in \mathsf{Gal}(\mathbb{Q}(\zeta_m)/\mathbb{Q})} \mspace{-20mu}\cn_G^{(m,n)},\qquad \cn_G^0\equiv \cn_G^{(1,1)}
\ee
where $\mathbf{C}(\mathfrak{g})$ is the \emph{set} (no repetitions!) of dual Coxeter labels\footnote{\ For simply-laced algebras
the dual Coxeter labels coincide with the Coxeter labels.} of the affine Dynkin graph of $\widetilde{\mathfrak{g}}$ and $n\in (\Z/m\Z)^\times\simeq \mathsf{Gal}(\mathbb{Q}(\zeta_m)/\mathbb{Q})$ is a choice of  a primitive $m$-root of unity.
The connections on the component $\cn_\mathfrak{g}^{m,n}$ are characterized by the value of
the Chern-Simons invariant\footnote{\ $h^\vee$ is the dual Coxeter number of $\mathfrak{g}$.}
\be\label{uuuuyyy12}
\int_{T^3}\omega^{(3)}\equiv \frac{1}{16\pi^2 h^\vee}\int_{T^3}\mathrm{tr}\!\left(A\,dA+\frac{2}{3}\,A^3\right)=\frac{n}{m}\bmod1
\ee

\begin{rem}\label{yyyyttt2} In the statement of \textbf{Theorem 1.1} of \cite{oguiso2} there is no mention of the choice of primitive $m$-th root.
However from the proofs and explicit constructions in \cite{oguiso2} one sees that one needs to choose a primitive root: see e.g.\! their eq.\eqref{seeeg}. Different choices lead to distinct multiple fibers with the same characteristic cycle type.
\end{rem}

\subsubsection{(Partially) frozen singularities in M-theory}
Partially frozen singularities in M-theory have been discussed in \cite{MM1,MM2,F2,YYY}.
\medskip

An unfrozen singularity of the form $\C^2/\Gamma_\mathfrak{g}$ (where $\Gamma_\mathfrak{g}\subset SU(2)$ is the finite subgroup corresponding to the $ADE$ Lie algebra $\mathfrak{g}$) yields  at the singular locus a 7d 16-SUSY super-Yang-Mills theory with gauge algebra $\mathfrak{g}$. In addition to its unfrozen 
version, the singularity has (partially) frozen variants, still preserving 16 supercharges, characterized by a fractional  3-form flux emerging from the singularity
\be\label{nontrivialflux}
\int_{S^3/\Gamma_\mathfrak{g}} C^{(3)}= \frac{n}{m}\bmod1.
\ee
As in eq.\eqref{uuuuyyy12} $m$ is a dual Coxeter label on a node of the Dynkin graph of $\mathfrak{g}$
 and $n\in (\Z/m\Z)^\times$.  $m$ is the \emph{order} of the singularity. The unfrozen singularity has order $1$.
 The flux is fractional iff $m\geq2$, i.e.\! if $m$ is the Coxter label of a non-extension node.
  Only Lie algebras of type $DE$ lead to frozen singularities,  since all nodes in a $\widetilde{A}_k$ Dynkin graph are
 extension.
 
  \paragraph{The associated Lie algebra $\mathfrak{f}^m_g$.}
 The 7-brane of a partially frozen singularities has a gauge symmetry Lie algebra $\mathfrak{f}^m_\mathfrak{g}$ which is smaller than $\mathfrak{g}\equiv \mathfrak{f}_\mathfrak{g}^1$. The dual Coxeter labels of its affine extension $\widetilde{\mathfrak{f}}^m_\mathfrak{g}$
satisfy the following rule \cite{MM2}: let $\{m_{i_k}\}_k\subset\mathbf{C}(\mathfrak{g})$ be the list of 
dual Coxeter labels  of $\widetilde{\mathfrak{g}}$ which are divisible by the multiplicity $m$;
the dual Coxeter labels of $\widetilde{\mathfrak{f}}^m_g$ are the integers $\{m_{i_k}/m\}_k$. This rule uniquely determines $\mathfrak{f}^m_\mathfrak{g}$ except when $m=2$ and $\mathfrak{g}=E_6$ (resp.\! $\mathfrak{D}_{k+4}$) where the dual Coxeter labels of $\mathfrak{f}^2_\mathfrak{g}$
 are $(1,1,1)$ (resp.\! $1,\dots,1$)
consistent with both $A_2$ and $C_2$ (resp.\! $A_n$ and $C_n$).
The Lie algebras $\mathfrak{f}^m_\mathfrak{g}$ for $m\geq2$
are listed in table  \ref{table4} (cf.\! Table 1 in \cite{F2}).
The deformation space of a partially frozen singularity ($m\geq2$) satisfies
\be\label{kkkiu78e}
\dim_\C\!\left(\begin{smallmatrix}\textbf{partially frozen}\\ \textbf{singularity}\end{smallmatrix}\right)=\mathrm{rank}\,\mathfrak{f}^m_\mathfrak{g} <
\mathrm{rank}\,\mathfrak{g}=\dim_\C\!\left(\begin{smallmatrix}\textbf{unfrozen}\\ \textbf{singularity}\end{smallmatrix}\right).
\ee

 To each pair $(\mathfrak{g},m)$ ($m\geq2$) we associate the Kodaira fiber type $\mathbf{K}_\mathfrak{g}$
 dual to its extended Dynkin graph $\widetilde{\Gamma}_\mathfrak{g}$,
 whose local monodromy $\varrho_\mathfrak{g}$ can be read
 in tables \ref{table},\,\ref{table2},
and also the Kodaira fiber $\widetilde{\mathbf{K}}_\mathfrak{g}$ obtained from it by the \emph{base change} $z\mapsto z^m$ \cite{miranda} i.e.\! the Kodaira fiber with local monodromy $\varrho_\mathfrak{g}^m$ (cf.\! table 2 of \cite{F2}). The resulting Kodaira types are listed in table \ref{table4}.
 
 \begin{table}
 \begin{footnotesize}
$$
\begin{tabular}{c|ccccccccccc}\hline\hline
pair & $(D_{k+4},2)$ & $(E_6,2)$ & $(E_6,3)$ & $(E_7,2)$ & $(E_7,3)$ & $(E_7,4)$ & $(E_8,2)$ & $(E_8,3)$ & $(E_8,4)$ & $(E_8,5)$ & $(E_8,6)$\\
$\mathbf{K}_\mathfrak{g}$ & $\mathbf{I}_k^*$ & $\mathbf{IV}^*$ &  $\mathbf{IV}^*$ &$\mathbf{III}^*$ & $\mathbf{III}^*$ &
$\mathbf{III}^*$ & $\mathbf{II}^*$ & $\mathbf{II}^*$ &$\mathbf{II}^*$ &$\mathbf{II}^*$ &$\mathbf{II}^*$\\
$\widetilde{\mathbf{K}}_\mathfrak{g}$ & $\mathbf{I}_{2k}$ & $\mathbf{IV}$ & $\mathbf{I}_0$ & $\mathbf{I}_0^*$ & $\mathbf{III}$ &
$\mathbf{I}_0$ & $\mathbf{IV}^*$ & $\mathbf{I}_0^*$ & $\mathbf{IV}$ & $\mathbf{II}$ & $\mathbf{I}_0$\\
$\mathfrak{f}^m_\mathfrak{g}$ & $C_k$ & $A_2$ & $0$ & $B_3$ & $A_1$ & $0$ & $F_4$ & $G_2$ & $A_1$ & $0$ & $0$\\\hline\hline
\end{tabular}
$$
\end{footnotesize}
\caption{\label{table4} In the first row we list the pairs $(\mathfrak{g},m)$
where $\mathfrak{g}$ is a simply-laced Lie algebra and $m$ a Coxeter label of a non-extension node of its Dynkin graph.
In the second row we list the Kodaira fiber type $\mathbf{K}_\mathfrak{g}$ associated to $\mathfrak{g}$. The third row gives the Kodaira fibers $\widetilde{\mathbf{K}}_\mathfrak{g}$ obtained from the ones in the second row by the base change $z\mapsto z^m$.  The forth row yields the gauge Lie algebra $\mathfrak{f}^m_\mathfrak{g}$ for the (partially) frozen M-theory singularity associated to the pair $(\mathfrak{g},m)$.}
\end{table}

\medskip

Comparing tables \ref{table4} and \eqref{multiple}, and bearing in mind that the explicit construction of the 
multiple fibers in \S.\,\ref{s:multiple} depends on a choice of a primitive $m$-root of unity $\zeta_m$,
we get:

\begin{fact}\label{conc} There is a one-to-one correspondence between Lagrangian multiple fibers in rank $r\geq2$
 (with trivial HO invariant in the
$\mathbf{I}_{2n}$ case) and (partially) frozen singularities in M-theory. The correspondence is as in the following table:
{\rm\be\label{corrr}
\begin{tabular}{r@{\quad }|@{\quad }l}
 Lagrangian singular fibers & M-theory singularities\\\hline\hline
fiber multiplicity $m$ & order $m$ of the singularity\\
monodromy $\varrho_\alpha$ & monodromy $\varrho_\mathfrak{g}$\\
characteristic cycle type & Kodaira type $\widetilde{\mathbf{K}}_\mathfrak{g}$\\
choice of primitive root $\zeta_m$ & $n\in (\Z/m\Z)^\times \simeq \mathsf{Gal}(\mathbb{Q}(e^{2\pi i/m})/\mathbb{Q})$\\
\end{tabular}
\ee }
\end{fact}

\begin{rem}
The correspondence \eqref{corrr} extends to \emph{simple} fibers (with trivial HO invariant), $m=1$,
of types different from $\mathbf{II}$, $\mathbf{III}$, $\mathbf{IV}$. That these three types should be excluded
looks physically natural since it reflects the QFT \emph{``fiber selection rule''} stated in the \textbf{Fact} of \S.\,3.3.1 in  \cite{C1}
which is the geometric counterpart to
 the \emph{``no dangerous irrelevant operator''} property (a.k.a.\! \emph{``safely irrelevant conjecture''})
 argued on physical grounds in
\cite{M1,M3,M4,M5,M6}. This observation may be seen as a geometric ``proof'' of the physical conjecture. 
\end{rem}

In the correspondence
\be
\left[\begin{smallmatrix}\textbf{\emph{simple} fibers $\neq \mathbf{II},\,\mathbf{III},\,\mathbf{IV}$ in Special\phantom{m}}\\
\textbf{Geometry (with trivial HO invariant)}\end{smallmatrix}\right]\longleftrightarrow
\left[\begin{smallmatrix}\textbf{\emph{unfrozen} singularities}\\ \textbf{in M-theory\phantom{mmmmn}}\end{smallmatrix}\right]
\ee
 the Lie algebra $\mathfrak{g}$ which is the gauge Lie algebra in M-theory becomes the flavor Lie algebra
 in the $\cn=2$ SCFT side. We shall see below using string dualities that this rule extends to the \emph{multiple} fiber as well, which then carry
 a flavor Lie algebra $\mathfrak{f}^m_\mathfrak{g}$ equal to the gauge algebra of the corresponding partially frozen M-theory singularity. 
 This also entails that the singularity is (partially) frozen on both sides of the correspondence, since the allowed deformations are related, respectively, to the flavor and gauge Lie algebras
 via eq.\eqref{kkkiu78e}.
 Comparing tables \ref{equuua},\,\ref{equuua2} and \ref{table4} we confirm the expectation
 \be
 \dim_\C\cm \equiv \mathrm{rank}\,\mathfrak{f}^m_\mathfrak{g},
 \ee
that is,
\begin{fact}\label{deffff} The deformation space $\cm$ of a multiple Lagrangian fiber {\rm (cf.\, \S.\,\ref{defspaces})} coincides with the deformation space of
the M-theory singularity with the same pair $(\mathfrak{g},m)$; in particular both have dimension $\mathrm{rank}\,\mathfrak{f}^m_\mathfrak{g}$.
\end{fact}

\begin{rem} A fiber with monodromy type $\mathbf{I}_b^*$ may be both double and have a covering
fiber with fiber cycle of type $\mathbf{I}_{2b}$ but characteristic cycle of type $\mathbf{I}_{2bm}$ for any $m\in\mathbb{N}\cup\infty$. In other words the ``freezing'' mechanisms described in \S.\,\ref{s:heuristics} and in the present section can operate \emph{simultaneously}: see \textbf{Example 6.3} in \cite{oguiso2}. This lead to a much large 
zoo of partially frozen singular Lagrangian fibers all with the 
 monodromy type $\mathbf{I}^*_b$. This opens the possibility of new phenomena which should be duly explored.
\end{rem}

\subsubsection{Duality with F-Theory} 
The identifications in \textbf{Facts \ref{conc},\,\ref{deffff}}
are no coincidence. They follow from the duality between M- and F-theory. 
We consider M-theory on the geometry
\be\label{kkkk4}
\ck \times S^1\times \R^2 \times (\R^{2,1}\times S^1),
\ee
where $\ck\to \mathbb{P}^1$ is an elliptic K3 and the radius of the second $S^1$
is parametrically large.
$\ck\to \mathbb{P}^1$ is a Lagrangian fibration.
 We suppose that the fiber $\ck_p$ over $p\in\mathbb{P}^1$ contains a $\C^2/\Gamma_\mathfrak{g}$ singularity. We consider a small disk $\Delta\subset\mathbb{P}^1$ centered at $p$, with complex coordinate $w$, and focus on the local Lagrangian fibration over it,
 $\ck|_{\Delta}\to \Delta$. The central fiber has Kodaira type $\mathbf{K}_\mathfrak{g}$,
 and the local monodromy around the origin in $\Delta$ of the elliptic fibration  is $\varrho_\mathfrak{g}$.  The local geometry is then
 \be\label{thisgeee}
 \ck|_{\Delta}\times S^1\times \R^2 \times (\R^{2,1}\times S^1).
 \ee
 We also assume that the singularity is (partially) frozen, i.e.\! we have a fractional 3-form flux as in equation \eqref{nontrivialflux}. 
 
At a generic point $0\neq w\in\Delta$ we can reduce to type IIA along one $S^1$ in the elliptic fiber $\ck_w\simeq (S^1)^2$,
 and then take the $T$-dual of the other $S^1$ in the fiber. In absence of 3-form flux we get the type IIB geometry \cite{F2,YYY}
 \be\label{xxxxyq12}
 \Delta\times S^1\times S^1\times \R^2\times(\R^{2,1}\times S^1)\simeq \Delta\times E\times \C\times(\R^{2,1}\times S^1)
 \ee
 where $E\simeq (S^1)^2$ is a fixed elliptic curve with complex coordinate $z$. 
Since the axio-dilaton $\tau$ is not constant, this situation is more appropriately described by 
a F-theory geometry: its 12-dimensional elliptically-fibered space is
 \be
 \ck|_\Delta\times E\times \C\times (\R^{2,1}\times S^1), 
 \ee
 and we have a 7-brane in the 10d base
  at $w = 0$ which is wrapped on $E\times \C\times(\R^{2,1}\times S^1)$. 
 
Now we switch on a non-trivial fractional 3-form flux \eqref{nontrivialflux} in this geometry.
The net effect \cite{F2} is that now
the monodromy action on the F-theory elliptic fiber as $w\to e^{2\pi i}w$
 is accompanied by a shift by $2\pi \breve{n} R^\prime/m$ of the $T$-dualized circle
(the first $S^1$ in the \textsc{lhs} of \eqref{xxxxyq12} of radius $R^\prime$) where
$\breve{n}$ is the inverse of $n$ in $(\Z/m\Z)^\times$
\be
\breve{n}\,n=1\bmod m.
\ee
In other words, the monodromy acts on the complex coordinate of elliptic curve
$E$
by a shift $z\mapsto z+p_m$ where $p_m\in E$ is a certain primitive\footnote{\ I.e.\! $m$ is the smallest positive integer so that $m p_m=0\in E$.} $m$-torsion point.

We probe this F-theory configuration with
 a D3 brane wrapped on the last factor space $(\R^{2,1}\times S^1)$;
 its image in the other factor spaces is a point.
 
To study this geometry it is convenient to introduce 
a covering disk with coordinate $t$
\be
 f\colon \widetilde{\Delta}\to\Delta,\qquad
 t\mapsto w=t^m.
 \ee
The pulled-back elliptic fibration $\widetilde{\ck}|_\Delta\to\widetilde{\Delta}$
has now monodromy $\varrho_\mathfrak{g}^m$ and Weierstrass equation
\be
y^2=P_0(x,t)\quad \text{with}\quad P_0(\zeta_m^2x,\zeta_m)=\zeta_m^6\,P_0(x,t).
\ee
Before introducing the D3 probe, our set-up is a 16-\textsc{susy} F-theory configuration over the 12d 
elliptic fibration 
\be\label{uuu712z}
(\widetilde{\ck}|\times E)/\Z_m\times \C\times (\R^{2,1}\times S^1)\to (\widetilde{\Delta}\times E)/\Z_m\times \C\times (\R^{2,1}\times S^1),
\ee
where $\Z_m$ acts as \cite{F2}
\be
\Z_m \ni k \colon \widetilde{\ck}|\times E\to \widetilde{\Delta}\times E,
\qquad k\colon (x,y,t,z)\mapsto (\zeta_m^{2k}x,\zeta_m^{3k} y,\zeta_m^kt, z+k p_m). 
\ee
with $p_m\in E$ a $m$-torsion point: the group $(\Z/m\Z)^\times$ (i.e.\! the $n$ in eq.\eqref{nontrivialflux}) acts by $p_m\mapsto \breve{n}\,p_m$. 
This statement is easily checked by taking the $T$-dual of both circles in the elliptic fiber in eq.\eqref{thisgeee}
and considering the extra $S^1$ factor in that equation as the M-theory circle, see \S.\,2 of \cite{F2}.

The 12-dimensional total space of the F-theory geometry is
\be\label{uuuuuu231}
\boldsymbol{Y}=(\widetilde{\ck}|\times E)/\Z_m\times \C\times (\R^{2,1}\times S^1)\equiv \boldsymbol{X}\times(\R^{2,1}\times S^1).
\ee
We note that $\boldsymbol{X}$ is a complex manifold of dimension 4
which carries a nowhere vanishing holomorphic 2-form 
\be\label{pppoooiq}
\Omega=\omega+ dz\wedge ds
\ee
where $\omega$ is the symplectic form on the local 2-CY $\widetilde{\ck}|$ and $z$ (resp.\! $s$) is the coordinate on
$E$ (resp.\! $\C$).
We conclude that the fibration $\boldsymbol{\pi}\colon\boldsymbol{X}\to \Delta\times \C$ is a rank-2 (germ) special geometry in its own right. By comparison with the Hwang-Oguiso construction in \S.\,\ref{s:explicit} we see that its central fiber is a multiple fiber of multiplicity $m$ with characteristic cycle of Kodaira type $\widetilde{\mathbf{K}}_\mathfrak{g}$.

\subsubsection{Effective theory on the D3 probe: unfrozen singularity}

 On the world-volume of the D3 probe there is a 4d 8-\textsc{susy} QFT whose flavor group contains the gauge group of the SYM theory living on the 7-brane.  This 4d QFT is compactified down to 3d on a \emph{large} circle $S^1$.  The circle compactification  allows us to describe the IR physics of the probe in both the languages of 4d $\cn=2$ and 3d $\cn=4$ supersymmetry.
 
We  consider first the case of an unfrozen singularity ($m=1$). 
The light 3-3 string states yield the d.o.f.\! of a 4d $\cn=4$ vector-multiplet, that is, a 4d $\cn=2$ vector-multiplet plus
a neutral $\cn=2$ hypermultiplet: from the viewpoint of 3d $\cn=4$ \textsc{susy} they correspond respectively to a twisted-hypermultiplet and a hypermultiplet which combine into a supermultiplet of 3d $\cn=8$ with R-symmetry algebra $\mathfrak{spin}(8)$ which from the 3d $\cn=4$ viewpoint yields
a $\mathsf{Spin}(4)$ R-symmetry and a $Sp(1)\times Sp(1)\simeq \mathsf{Spin}(4)$ flavor group. The v.e.v.\! of the scalars in the two 4d supermultiplets describe the position of the probe:
the vector-multiplet in the directions orthogonal to the 7-brane (whose support is $D_\alpha$)
and the hypermultiplet in the parallel ones. It is well known (see \cite{probing,Hamada:2021bbz})
that the 3d twisted-hypermultiplet scalars describe the position of the D3 brane in the full
elliptic hyperK\"ahler manifold $\ck$. The 3-7 and 7-3 strings produce interacting d.o.f.
coupled to the vector multiplet living on the probe. This interacting sector has a flavor Lie algebra
$\mathfrak{g}$ equal to the gauge algebra of the theory living on the 7-brane.  
Thus the IR effective theory on the D3 probe is an interacting rank-1 4d $\cn=2$
 with flavor algebra $\mathfrak{g}$ plus an IR-decoupled free hypermultiplet described by a 8-\textsc{susy} $\sigma$-model with target the factor space in \eqref{uuu712z} (for $m=1$) which is parallel to the 7-brane and normal to the probe,
 i.e.\! the flat hyperK\"ahler manifold $\ch\equiv E\times \C$ equipped with its natural holomorphic symplectic form
 $ds\wedge dz$. This sector has its own flavor algebra $\mathfrak{sp}(1)$.
 
 In the 3d description the IR theory is a $\cn=4$ $\sigma$-model with 
 target  space the product two hyperK\"ahler manifolds each of quaternionic dimension 1
 \be\label{juqyas}
\ck\times \ch
 \ee
The factor $\ch$ is the target space of the hypermultiplet scalars (the same manifold as in the 4d description),
while the elliptic 2-CY $\ck$ is the target space for the twisted-hypermultiplets.
In the standard conventions of 3d $\cn=4$ \textsc{susy} \cite{Cecotti:2015wqa}
the orientations of the two factor spaces are chosen so that
the Riemann tensor of $\ch$ is selfdual while the Riemann tensor of $\ck$ is anti-selfdual. Geometrically it may be more natural to
 redefine the orientations so that the product space is a single hyperK\"ahler of double dimension
 with the holomorphic symplectic form   \eqref{pppoooiq}.

Before turning to the $m>1$ case we digress on fine points about supersymmetry.

\subsubsection{Subtleties with SUSY factorization}
It is usually stated, and used to deduce countless \textsc{susy} non-renormalization theorems,\footnote{\ See e.g.\! \cite{mirrorbook} for basic examples.} that when a 
\textsc{susy} model contains fields transforming in different supermultiplets, the
scalars take value in a Riemannian product
\be\label{ju1234}
M_1\times M_2\times\cdots\times M_\ell,
\ee  
 where $M_i$ is the target space of the scalars belonging to supermultiplets of the $i$-th type.
This statement applies to chiral versus twisted-chiral in 2d $(2,2)$ \textsc{susy} \cite{mirrorbook}, or
 to hypers vs.\! twisted-hypers in 3d $\cn=4$ \textsc{susy} \cite{cmap}, or vector-multiplets vs.\! hypermultiplets in 4d $\cn=2$ \textsc{susy} \cite{cmap}, etc.
 However the traditional argument   technically proves only a \emph{weaker} statement \cite{Cecotti:2015wqa}:
\begin{ped}
If the scalars' target space $M$ of a \textsc{susy} field theory is a \emph{complete,} \emph{simply-connected,} Riemannian manifold,
 then $M$ is a metric product
 as in eq.\eqref{ju1234}.
 \end{ped}
\noindent There are two clauses in the statement.  The target space of a \textsc{susy} effective theory with $\leq8$ supercharges is usually non-complete: typically there are singularities at finite-distance where some additional d.o.f.\! get light.
A more sophisticated analysis shows that \emph{when we have 8 supercharges}
 we can get rid of the completeness hypothesis:\footnote{\ In particular the validity of the non-renormalization theorems is not spoiled by the subtleties discussed in this subsection.}
 the statement remains true (with qualifications) in presence of the mild singularities expected on physical grounds\footnote{\ At least when the IR effective theory \emph{do admit} a UV completion.}. 
The assumption that the scalar space is
 \emph{simply-connected} is instead essential for the factorization property to hold, as the (counter)examples below will show. 

We can always reduce ourselves to the simply-connected case by replacing the scalar manifold $M$
by a smooth simply-connected cover\footnote{\ A smooth simply-connected cover is a finite cover of the universal cover which smooths out the orbifold singularities of the topological universal cover.} $\widetilde{M}$ so that $M\equiv \widetilde{M}/\Gamma$. Thus all \textsc{susy} model is obtained from one where the scalars take value in the product manifold $\widetilde{M}$ by gauging a discrete symmetry
$\Gamma$ which acts on both factors simultaneously. In other words: the factorization holds at the level of the \emph{local physics,} but it may be spoiled by global topological effects which couple, say,
the hypermultiplets and the twisted-hypermultiplets of a 3d $\cn=4$ $\sigma$-model.

The gauging of the discrete group $\Gamma$ preserves all supersymmetries iff it acts
on each factor space by automorphisms of the relevant geometric structures. For instance, when the covering space
 is the product of two hyperK\"ahler manifolds $\ck\times\ch\equiv \widetilde{M}$ as in eq.\eqref{juqyas},
$\Gamma$ should act by simultaneous symplectic holomorphic isometries
 on both $\ck$ and $\ch$. Isometries of this kind exist,
so we can indeed have 3d $\cn=4$ $\sigma$-models with globally non-factorized scalar target spaces.
 When the 3d $\cn=4$ $\sigma$-model arise as the low-energy limit of a 4d $\cn=2$ QFT compactified on $S^1$ one says that $\ck$ is the ``3d Coulomb branch'' and $\ch$ the ``Higgs branch''.
Both branches are well-defined only up to a cover.
There are strong reasons to believe  that in a UV-complete theory
the Coulomb (resp.\! Higgs) branch is well-defined up to \emph{finite} covers \cite{Cecotti:2020rjq}.
\medskip

Now suppose  that the factor $\ch$ is smooth and flat. Before the discrete gauging, the hypermultiplets form a free sector which decouples from the rest\footnote{\ In the IR, hence in the full theory when it is a SCFT.}. After the gauging of $\Gamma$, the hypermultiplets still look like a free decoupled sector at the level of the \emph{local physics}, but \emph{globally} they are still coupled to the interacting sector by a topological feature.
This is another version of the mechanism discussed in \S.\ref{s:2language}: 

\begin{phm} The \textbf{local physics} contains a free decoupled subsector which however in the full theory
is still coupled by subtle global topological effects.
\end{phm}
   The difference between the present case and the one in \S.\ref{s:2language} is that there the would-be free sector was a set of $(r-1)$ vector-multiplets whereas now are hypermultiplets. The physical principle in action is nevertheless the same, and mathematically they both arises from the fact that the Kodaira type of the Hwang-Oguiso characteristic cycle $\boldsymbol{\Psi}$ does not match the type of the local monodromy.
   
 \subsubsection{Effective theory on the D3 probe: (partially) frozen singularity}  

We return to our D3 brane probing the F-theory geometry \eqref{uuuuuu231} which is dual to a frozen M-theory singularity.
We consider its 3d $\cn=4$ IR effective theory.
From the geometry it is obvious that this effective theory is obtained by gauging a $\pi_1(\boldsymbol{X})\equiv \Z_m$
symmetry of a system containing a twisted-hypermultiplet taking value in $\ck$ and a hypermultiplet valued in $\ch$.
Since $\ch$ is locally flat, the hypermultiplet is \emph{locally} free yet globally coupled to the twisted hypermultiplet
by the $\Z_m$ discrete gauging. A peculiarity of this 3d $\cn=4$ model is that also the hyperK\"ahler factor $\ch$ is elliptically fibered
with Lagrangian fibers and so is a (free) special geometry in its own right.

Since both factor spaces of the universal cover $\widetilde{\boldsymbol{X}}$ are elliptic fibered, 
we may consider the projection of $\boldsymbol{X}$ on a 2-dimensional complex base $\widetilde{\Delta}/\Z_m\times \C$
of coordinates $w=t^m\!,$ $s$
\be\label{kkkk5554z}
\boldsymbol{\varpi}\colon \boldsymbol{X}\equiv (\widetilde{\ck}|\times E)/\Z_m\times \C\longrightarrow  \widetilde{\Delta}/\Z_m\times \C\simeq \Delta\times \C\subset \C^2,
\ee
which is a Lagrangian fibration whose generic fibers are Abelian varieties of dimension 2, that is,
$\boldsymbol{\varpi}$ is a rank-$2$ special geometry. In fact $\boldsymbol{\varpi}$ is a ``no-section'' special geometry (in the sense of \S.\ref{s:preliminary}) with a central fiber of multiplicity $m$ equal to the order of the original M-theory (partially) frozen singularity.
This conclusion is not in contradiction with the physical fact that the Seiberg-Witten special geometry along the Coulomb
branch must be of the ``section'' kind (cf.\! \S.\,\ref{s:preliminary}) since the QFT on the probe \emph{has rank 1} and the two-dimensional
base of the Lagrangian fibration \eqref{kkkk5554z} has \emph{not} the physical interpretation of being its Coulomb branch. Moreover the statement in \S.\,\ref{s:preliminary} that the Seiberg-Witten geometry has a zero section
holds in the same sense in which the notion of ``Coulomb branch'' is defined, that is, only modulo finite covers.
 
Comparing with the Hwang-Oguiso
construction in \S.\,\ref{s:multiple}, we see that the rank-2 special geometry \eqref{kkkk5554z}
along the discriminant divisor $t=0$  has precisely a singular fiber of multiplicity $m$ with characteristic cycle $\boldsymbol{\Psi}$ of type $\widetilde{\mathbf{K}}_\mathfrak{g}$. 
This explains the correspondence in \textbf{Fact \ref{conc}}:
 duality with $F$-theory in presence of a D3 probe
 transforms a 
(partially) frozen M-theory singularity into a multiple fiber of the ``no-section'' 
rank-2 special geometry which describes the topologically-coupled 
3d twisted-hyper/hyper system living on the D3 probe wrapped on $S^1$.

The gauge algebra $\mathfrak{f}^m_\mathfrak{g}$ of the theory living on the frozen 7-brane (cf.\! table \ref{table4})
becomes part of the flavor symmetry of the theory on the probe.
The deformation space of the geometry is the space constructed in
\S.\,\ref{defspaces} of dimension $\mathrm{rank}\,\mathfrak{f}^m_\mathfrak{g}$.

\section{Implications for the $\cn=2$ SCFT classification program}\label{s:classs}

There is an ongoing program to classify all possible 4d $\cn=2$ SCFT
by classifying the corresponding scale-invariant special geometries: see \cite{M1,M2,M3,M4,M5,M6,C1,C2} for a partial list of references.
The present discussion has several important implications for that program.
Here we limit ourselves to the most obvious ones: finer points will be discussed elsewhere.

\subsection{What are we supposed to classify?}\label{s:what}

Looking, say, at the list of know rank-1 SCFTs, table 1 of \cite{M1},
one gets the impression that to a given scale-invariant special geometry 
there correspond several distinct SCFTs; in other words
it seems that the geometry is not powerful enough to detect all the physical subtleties of the SCFTs. 
In particular the ``same'' special geometry describes SCFTs whose flavor groups have different ranks, so that their mass-deformation spaces have different dimensions. The traditional strategy works in two steps: \textbf{(1)} first one classifies all  scale-invariant special geometries in rank $r$, and
then \textbf{(2)} for each special geometry in the list one
one construct the (finitely many) physical SCFTs associated to it.
The second step is non-geometric, which means pretty hard. 
\medskip

Can we do better?
\medskip

The results of the present paper suggest a refined geometrical approach to the classification program.
Preliminarily we stress that ``to classify SCFTs in rank-$r$''
actually means to classify all ``minimal'' such theories.\footnote{\ Our language is modeled on the birational classification of compact complex surfaces.} Given a rank-$r$ SCFT we can get a ``non-minimal'' one by adding a free hypermultiplet or a free vector-multiplet. 
To reduce to the ``minimal'' SCFT we have to decouple all hyper- and vector-multiplets
which can be eliminated without introducing pathologies in the physics. In particular the 
process should preserve UV completeness. The reduction to ``minimal'' SCFT is a subtle operation: as we saw in the previous sections, the extra d.o.f.\!
may be decoupled at the level of local physics, yet globally coupled. The theory obtained by omitting the locally-decoupled d.o.f.\!  is still consistent in the UV (since the UV is sensitive only to the local dynamics), and we can \emph{define}
a ``minimal'' SCFT by extending the reduced consistent  UV description to all scales. 
But we pay a price for this: the standard-sense Seiberg-Witten geometry of the resulting ``minimal'' SCFT does not
contain all the information of the original special geometry, in particular we lose track of the global aspects associated to the topological couplings.
To retain all the informations encoded in the geometry we need to classify all ``non-minimal'' geometries, which describe also additional degrees of freedom, 
and only at the end reduce to the ``minimal'' SCFTs they describe.
The ``non-minimal'' geometries may not be ``section'' special geometries since they have not the
physical interpretation of standard Seiberg-Witten geometries.

For the purpose of this section it is enough to consider the case where 
the extra ``non-minimal'' d.o.f.\! are hypermultiplets.    
A necessary condition for a hypermultiplet to be eliminable is that it is light everywhere along the Coulomb branch $\mathscr{C}$. Locally at a generic vacuum such a hypermultiplet looks just free. The crucial question is whether it can be eliminated without losing the UV completeness.
To get the ``minimal'' SCFT we have to decouple all everywhere light hypers
whose elimination do not introduce singularities. At the end we remain with a number $h$
of everywhere light hypers which are part of the ``minimal'' SCFT.  

\begin{exe} Consider the class $\cs$ theory obtained by the 6d $(2,0)$ theory of type $A_1$
on a punctured torus. It is $\cn=2$ SYM with $G=SU(2)$
and two hypermultiplets in the $\boldsymbol{1}\oplus\boldsymbol{3}$ of the gauge group (cf.\! \S.\,4.8 of \cite{Alim:2011kw}). This
model has two everywhere light hypermultiplets: one from the $\boldsymbol{1}$ and
one from the Cartan subalgebra in the $\boldsymbol{3}$.  The first one is eliminable, the second 
one not. This is pretty obvious from the Lagrangian description of the model. Geometrically it follows from the fact that the second hyper
changes sign under the action of the $SU(2)$ Weyl group so it takes values in the singular space $\C^2/\Z_2$. Alternatively: the global field describing this everywhere light hyper has dimension 2 not 1 and hence is not a free subsector. We conclude that this example has $h=1$. 
\end{exe}

When $h\geq1$ in the IR we have an emergent symmetry $F_\text{em}\subseteq\mathsf{Sp}(h)$
which acts effectively and (typically) irreducibly on the light hypers. Since the theory is conformal, it ``looks the same''
at all scales, and at least the \emph{rank} of the quotient of the flavor group which acts effectively
on the everywhere light hypers in the full ``minimal'' theory should agree with $\mathrm{rank}\,F_\text{em}$.

\subsection{Rank-$1$ \emph{geometric} classification reloaded}\label{s:revisited}

As an example of the afore mentioned refined 
geometric classifying strategy, we revisit the classification of $\cn=2$ SCFTs of rank-$1$. This classification has been worked out in \cite{M1,M3,M4,M5,M6,C1,C2}.

According to the previous subsection we should not limit ourselves to consideration of
Seiberg-Witten geometries over Coulomb branches of complex dimension 1: 
 we have to 
study all ``non-section'' special geometries $\mathscr{Y}\to\mathscr{B}$,
where $\dim\mathscr{B}$ may be larger than 1, which have a discrete cover of the form $\ck\times \ch$
with $\ck$ of rank-1 and a singular fiber in codimension-$1$ with the appropriate monodromy class.

The conjugacy class of the monodromy $\varrho_\mathfrak{g}$ around the codimension-$1$ singularity
 determines the physical Coulomb branch dimension $\Delta$.
Only the semisimple conjugacy classes
yield SCFTs. The list of dimensions $\Delta$ for each torsion conjugacy classes in $SL(2,\Z)$ is given in  table \ref{rrrowwwa}: we label each class by both its Kodaira type and the corresponding Lie algebra $\mathfrak{g}$. 
The classification of the singular fibers leads to the

\begin{bid}
The refined geometric classification of rank-1 ``minimal'' SCFTs with a number
of supercharges of the form $2^{k+3}$ ($k\geq0$) requires \textbf{as a first step} 
to  classify  
all scale-invariant \textbf{rank-2} special geometries (with or without section)
with an irreducible \textbf{smooth} discriminant
whose  local monodromy is semisimple.
\end{bid}

\begin{rem}\label{jutree} In the language of \cite{M2} we require the rank-2 special geometry to have a single
unknotted discriminant. When the geometry is of the Seiberg-Witten (SW) type (``section'') this implies that special geometry is a product
of a scale-invariant rank-1 geometry and a free one  \cite{M2}. The non-free factor is the SW special geometry of the ``minimal'' rank-1 SCFTs. The ambient rank-2 geometry has a deformation space of dimension $\mathrm{rank}\,\mathfrak{g}$ and hence the SCFT has a flavor Lie algebra $\mathfrak{g}$.  
A part for the dummy free factor, these are the usual rank-1 geometries.
In addition we also have the rank-2 scale-invariant special geometries with a single unknotted discriminant which are \emph{not of the SW type}.
\end{rem}

\begin{rem} While the study of the rank-2 geometries in \textbf{Suggestion} may be just a first step in the rank-1 SCFT geometric classification it is certainly a very substantial step.
\end{rem}

The singular fiber\footnote{\ All fibers over the discriminant are isomorphic by the $\C^\times$ action.} over the discriminant may be unfrozen, partially frozen, or fully frozen ($\equiv$ rigid). 
We claim that the ``minimal'' SCFT described by a special geometry as in {\bf Suggestion}
has $h=0$ unless its singularity is \emph{fully} frozen. The idea is that when the multiple fiber is not rigid we can deform it into several simple
singular fibers which have $h=0$. When $h=0$ the flavor Lie algebra
is $\mathfrak{f}_\mathfrak{g}^m$, in particular, $\mathfrak{g}$ for the unfrozen ($m=1$)  case cf.\!
\textbf{Remark \ref{jutree}}.

The $m=1$ SCFTs for the various values of $\Delta$ form the 
Minahan-Nemeschancky (MN) series: the models and their flavor algebras $\mathfrak{g}$ are listed in 
table \ref{rrrowwwa}. In the table AD $G$ stands for the Argyres-Douglas model of type $G$ \cite{AD1,AD2,AD3,Cecotti:2010fi}
and MN $F$ for the MN  model with flavor group $F$ \cite{MN1,MN2}. $SU(2)$ $N_f=4$
stands for the $\cn=2$ SQCD with 4 hypermultiplet flavors in the fundamental.
%
%

\begin{table}
\begin{tabular}{c|ccccccc}
$\begin{smallmatrix}\textbf{monodromy}\\
\textbf{class} \end{smallmatrix}$ & $\mathbf{II}$ & $\mathbf{III}$ & $\mathbf{IV}$ & $\mathbf{I}_0^*$ & $\mathbf{IV}^*$ & $\mathbf{III}^*$ & $\mathbf{II}^*$\\
$\mathfrak{g}$ & $0$ & $A_1$ & $A_2$ & $D_4$ & $E_6$ & $E_7$ & $E_8$\\\hline
$\Delta$ & $\frac{6}{5}$ & $\frac{4}{3}$ & $\frac{3}{2}$ & $2$ & $3$ & $4$ & $6$\\
$\begin{smallmatrix}\textbf{MN series}\end{smallmatrix}$ & AD $A_2$ & AD $A_3$ & AD $D_4$ &
$SU(2)$ $N_f=4$ & MN $E_6$ & MN $E_6$& MN $E_6$\\
\end{tabular}
\caption{\label{rrrowwwa} The Coulomb dimensions $\Delta$ for each semisimple monodromy conjugacy class.
The last row gives the rank-$1$ SCFTs with maximal flavor symmetry for each $\Delta$} 
\end{table}
\medskip
%

We now consider each $\Delta$ in turn.

\subsubsection{Dimensions $\Delta=\tfrac{6}{5},\;\tfrac{4}{3},\;\frac{3}{2}$}
There are no multiple fibers with monodromy types  $\mathbf{II}$, $\mathbf{III}$, $\mathbf{IV}$.
Hence when $\Delta$ is fractional we do not get any new SCFT apart for the three Argyres-Douglas models of types $A_2$, $A_3$ and $D_4$ with flavor algebra $0$, $A_1$ and $A_2$ respectively.  

\subsubsection{Dimension $\Delta=2$} 
$m=1$ yields $SU(2)$ SQCD with $N_f=4$.
The only multiple fiber has $m=2$.
 $\Delta=2$ implies that the SCFT has a Lagrangian formulation, so $(\Delta=2,m=2)$ should correspond to the only
other rank-1 Lagrangian SCFT, namely $\cn=4$ SYM with gauge group $SU(2)$  seen as a special instance of $\cn=2$ SCFT.
Since $\mathfrak{f}^2_{\mathfrak{so}(8)}=0$, we expect the flavor algebra to be $\mathfrak{sp}(h)$ with $h$
the number of everywhere light hypers. This indeed holds for  $SU(2)$ $\cn=4$ SYM with $h=1$.

To see that this identification is indeed correct, we look to the corresponding F-theory geometry
\eqref{uuuuuu231} in the decoupling limit where the size of $E$ goes to zero. We can
study this configuration by the method of images: we uplift $\boldsymbol{X}\equiv(\widetilde{\ck}|\times E)/\Z_2\times \C$ to its double cover $\hat{\boldsymbol{X}}\equiv(\widetilde{\ck}|\times E)\times \C$
  whose central fiber is now smooth, i.e.\! there is no 7-brane anymore.
The only degrees of freedom on the probe are the ones from the 3-3 strings.
However in the double cover now we have \emph{two} D3 probes,
at opposite positions $t$, $-t$ so that their center-of-mass vector-multiplet is frozen
while the center-of-mass hypermultiplet should be decoupled to reduce to the ``minimal'' model.
The world-volume theory on the probe is now $SU(2)$ $\cn=4$ SYM.
The $\Z_2$ deck symmetry of the cover $\hat{\boldsymbol{X}}\to\boldsymbol{X}$ interchanges the two D3's, that is, acts as the $SU(2)$ Weyl group which is part of the gauge symmetry.
The Cartan hypermultiplet is then not eliminable (cf. \textbf{Example} in \S.\,\ref{s:what})
and the ``minimal'' SCFT has $h=1$.

\subsubsection{Dimension $\Delta=3$} For $\Delta=3$ we have three multiplicities
$m=1,2,3$. 
$(\Delta=3,m=1)$ is MN $E_6$.

\subparagraph{The $(m=\Delta=3)$ geometry.} The four geometries with $m=\Delta$
behave all in the same way and produce the Argyres-Wittig models \cite{AW}. They have flavor groups
with $\mathrm{rank}\,F=\mathrm{rank}\,\mathfrak{g}-3$ of the form $F=Sp(h)\times L$
for $L$ of rank $\mathfrak{g}-3-h$ \cite{C2}.
The first geometry in this series is the $m=\Delta=2$ one which we analyzed above by the method of images.
Now we consider the geometry with $m=\Delta=3$ using the same covering trick.
We have a new issue: the corresponding F-theory geometry depends on a choice
of $n\in(\Z/3\Z)^\times$. \emph{A priori} we may get two distinct ``minimal'' SCFTs.
However this is not the case since for all $n$ we have the same cover $\hat{\boldsymbol{X}}\to \boldsymbol{X}$ with the same deck group.
The argument applies to all $m=\Delta$ geometries.

The triple cover $\hat{\boldsymbol{X}}$ of $\boldsymbol{X}$ has no 7-brane and {three}
D3 probes at position $t$, $\zeta_3 t$, $\zeta_3^2 t$. Taking the size of $E$ to zero, this can be seen as the geometry describing $SU(3)$ $\cn=4$ SYM.
In particular the
 $\Z_3$ deck group of the cover $\hat{\boldsymbol{X}}\to\boldsymbol{X}$ is generated by the Coxeter element $C\in\mathsf{Weyl}(SU(3))$.
The covering special geometry is the \emph{restriction} of the $SU(3)$ $\cn=4$ SW geometry
 to the locus $W$ in the $SU(3)$ Coulomb branch $\mathscr{C}\equiv\C^2$ (with coordinates
$(u_2,u_3)$ and $\Delta(u_a)=a$) where
\be
z^3+u_2\, z-u_3 z\equiv\prod_{k=0}^2 (z-\zeta_3^k\, t)\equiv z^3-t^3\quad\Rightarrow\quad
W\equiv\begin{cases}u_2=0\\ u_3=t^3\equiv w\end{cases}
\ee
We consider a point $u\equiv (0,w)\in W$ with $w$ generic. Since $W\not\subset \mathscr{D}$ the fiber $X_u$ is a smooth Abelian variety. 
The $U(1)_R$ transformation $e^{2\pi iR/3}$ is unbroken in the vacuum $u$: it fixes  $W$ pointwise while acting
on $T^*_u\mathscr{C}\equiv \C\, du_2\oplus \C\, du_3$ as $\mathsf{diag}(\zeta_3^2,1)$. From the isomorphism
$\Omega\colon \mathsf{Lie}(X_u)\to T^*_u\mspace{2mu}\mathscr{C}$, we conclude that 
$f\equiv-e^{2\pi iR/3}$ is an automorphism of $X_u$ with analytic representation
$\rho_a(f)=\mathsf{diag}(\zeta_6,\zeta_6^{-1})\in SL(2,\C)$. Using
\textbf{Lemma 13.4.3} of \cite{complexabelian}, we get\footnote{\ \label{oooo1}The $\cn=4$ special geometry is isotrivial \cite{Cecotti:2021ouq} so eq.\eqref{ju71234b} applies everywhere on the $SU(3)$ $\cn=4$ Coulomb branch.}
\be\label{ju71234b}
X_u\simeq E_{\zeta_3}\times E_{\zeta_3}
\ee
while the Hamiltonian vector $v_3$ with $\iota_{v_3}\mspace{1mu}\Omega=du_3\equiv dw$ spans the complex Lie algebra of the second
$E_{\zeta_3}$. We conclude that we have a special sub-geometry of rank-1,
whose Coulomb branch $W$ has coordinate $w$ of dimension $\Delta=3$,
with fiber $E_{\zeta_3}$. 

In the language of classical integrable systems, passing from the rank-2
special geometry of $SU(3)$ $\cn=4$ to the rank-1 $\Delta=3$ special \emph{sub}-geometry
is the symplectic
reduction (Liouvillian separation of variables) where we set the first conserved momentum, $u_2$, to zero and then quotient out the flow of the Hamiltonian vector $v_2$ with $\iota_{v_2}\mspace{1mu}\Omega=du_2$ which is the first $E_{\zeta_3}$ in \eqref{ju71234b}. We remain with a reduced scale-invariant rank-1
integrable system with over the momentum space $W$ with $\Delta=3$ and fiber $E_{\zeta_3}$. 

It remains to compute $h$ for this special sub-geometry. 
The 3-3 everywhere light hypers take value
in $\C^3/\langle C\rangle$.
$\langle C\rangle\simeq \Z_3$ acts on $\C^3$ via the regular representation. 
The trivial representation has multiplicity 1 
and may be decoupled without introducing singularities (it is the center-of-mass of the 3 images of the probe). The other two linear combinations correspond to non-trivial characters and so cannot be safely omitted. Hence $h=2$.
In more intuitive terms: the number of everywhere light hypers coincides with $h\equiv \mathrm{rank}\,G$ of the
ambient $G=SU(3)$ $\cn=4$ geometry. There is an Argyres-Wittig rank-1 SCFT
with $\Delta=3$, $h=2$ and $F=Sp(2)\times U(1)$ \cite{AW} with the properties predicted by the
$m=\Delta=3$ geometry.

\subparagraph{The $(\Delta=3,m=2)$ geometry.}
By the \textbf{Claim} we expect $h=0$ and flavor $\mathfrak{f}^2_{E_6}=\mathfrak{su}(3)$. 

\medskip

This geometric analysis yields all flavor symmetry groups for rank-1 $\Delta=3$ SCFTs listed
in table 1 of \cite{M1} (their models $\#19,\,20,\,21$) \emph{but} the last one ($\#22$) which has $F=U(1)$.
However this model has $\cn=3$ SUSY (12 supercharges) whereas it is clear from the F-theory duality that the present discussion refers only to geometries which represent models with a total number of supercharges which is a power of $2$
and divisible by 8. Thus $\#22$ cannot correspond to a Hwang-Oguiso multiple fiber and is not cover by the \textbf{Suggestion}. 

\subsubsection{Dimension $\Delta=4$} We have $m=1,2,3,4$. As before the cases $m=1,2,3$ are expected to have $h=0$ and to correspond to
SCFTs with flavor algebra respectively
\be
\mathfrak{f}^m_{E_7}=E_7,\quad \mathrm{Spin}(7),\quad SU(2)
\ee
 and $h=0$. This is the full list of flavor symmetries in known SCFTs with $\Delta=4$ and $h=0$. 
 
 The case $m=\Delta=4$ is again tackled with the method of images. The covering geometry
 has no 7-brane and is a symplectic reduction of the
 special geometry of $SU(4)$ $\cn=4$ SYM on the axis $W=\{u_2=u_3=0\}$
 parametrized by the coordinate $u_4=w$ of dimension 4.
 The automorphism $\exp(2\pi i R/4)$ of the fiber $X_u$ over the generic point $u\in W$ has
 analytic representation $\rho_a(e^{2\pi i R/4})=\mathsf{diag}(\zeta_4,\zeta_4^2,\zeta_4^{-1})$
 so that\footnote{\ Footnote \ref{oooo1} still applies.}
 \be\label{iiiison}
 X_u \sim E \times A_{\zeta_4} \simeq E\times E_{\zeta_4}\times E_{\zeta_4}
 \ee
 where $\sim$ stands for isogeny, $E$ is an elliptic curve, $A_{\zeta_4}$ is an Abelian surface
 with a $\Z_4$ automorphism of analytic representation $\mathsf{diag}(\zeta_4^k,\zeta_4^{-k})$,
 and $E_{\zeta_4}$ an elliptic curve with period $\tau=\zeta_4\equiv i$. The isogeny in \eqref{iiiison} follows from \textbf{Theorem 13.2.8} of \cite{complexabelian} and the  isomorphism from \textbf{Lemma 13.4.3} in the same book. The Hamiltonian vector $v_4$ with $\iota_{v_4}\Omega=du_4\equiv dw$
 generates the Lie group of the last elliptic factor, and the symplectic reduction
 on the locus $u_2=u_3=0$ produces a rank-1 special sub-geometry
 with $\Delta=4$.
 Again, we expect $h=\mathrm{rank}\,SU(4)=3$ and $F=Sp(3)\times L$ with $\mathrm{rank}\,L=1$.
This matches with the existence of a rank-1 $\Delta=4$ SCFT with $F=Sp(3)\times Sp(1)$ \cite{AW,M1}.
 
 We got the full list of flavor groups known to appear for rank-1 $\Delta=4$.
 But some of the groups may be realized by inequivalent SCFTs with (in particular) different $h$'s.

\subsubsection{Dimension $\Delta=6$} 
We have $m=1,2,3,4,5,6$. For $m=1,2,3,4$ we get special geometries with flavor group respectively
\be\label{ju76123}
E_8,\quad F_4,\quad G_2,\quad SU(2).
\ee 
Again eq.\eqref{ju76123} yields the \emph{exact list} of flavor groups known to be realized in rank-1 SCFTs with $\Delta=6$ and $h=0$ \cite{M1,M6}.

$(E_8,m=6)$ can be analyzed with the method of images. The 6-fold cover has no 7-brane,
and the generic fiber $X_u$ over the Coulomb branch axis $H=\{u_2=u_3=u_4=u_5=0\}$
is a smooth Abelian 5-fold with a $\Z_6$ automorphism with analytic representation
\be
\rho_a(e^{2\pi i R/6})=\mathsf{diag}(\zeta_6,\zeta_3,-1,\zeta_3^{-1},\zeta_6^{-1})
\ee
so that
\be
X_u\sim A_{\zeta_3}\times E\times A^\prime_{\zeta_6}\simeq E_{\zeta_3}\times E_{\zeta_3}\times E\times E_{\zeta_3}\times E_{\zeta_3}
\ee
and the Hamiltonian vector dual to $dw$ spans the Lie algebra of the last factor.
Again we get a special sub-geometry with $\Delta=6$ and $h=\mathrm{rank}\,SU(6)=5$.
Therefore we expect the geometry to have a flavor group $Sp(5)$. Indeed a Argyres-Wittig rank-1
$\Delta=6$ SCFT with these properties exists \cite{AW,M1}.

\subsubsection{The last multiple fiber $(E_8,m=5)$} 
This is the hard case. The application of the image method to
this geometry is not straightforward since its F-theory
5-fold covering  geometry still contains a non-trivial 7-brane of type $\mathbf{II}$
which is probed by five D3's. Now the characteristic cycle $\boldsymbol{\Psi}$
 is \emph{not} a smooth elliptic curve contrary to the $m=\Delta$ case.

The main issue is to compute the number $h$ of everywhere light hypermultiplets. 
Now it is not clear \emph{a priori} if 
$h$
is independent of $n\in(\Z/5\Z)^\times$ or not. 
We give an \emph{indirect} argument showing that the physics of the F-theory
configuration depends on $n$ in an essential way,
which is consistent with the idea that 
 $h$ may be a non-trivial function of $n$.
We return to the F-theory geometry \eqref{uuu712z} which (after Wick rotation) we rewrite in the form
\be\label{xxxxx65123}
(\widetilde{\ck}\times S^1)/\Z_5\times S^1\times \C\times \R^{3}\times S^1\to (\widetilde{\Delta}\times S^1)/\Z_5\times S^1\times\C\times \R^{3}\times S^1,
\ee
To show that the physics of this system depends in an essential way from $n\in (\Z/5\Z)^\times$ we scan the geometry \eqref{xxxxx65123} with a \emph{different} probe:  a \emph{longitudinal} D3 wrapped on $\R^{3}$ times the \emph{first} $S^1$ instead of the \emph{last} $S^1$, so now the shift of the circle by $2\pi \breve{n}R^\prime/5$ is a translation along the probe
world-volume instead of a translation of the probe in a transverse direction. Following ref.\! \cite{Cecotti:2010fi}
we may topologically twist this new set-up by replacing $\R^3$ with the Melvin cigar $M\mspace{-1mu}C_q$. The partition function of the topologically twisted $\cn=2$
living on the loingitudinal D3 probe is \cite{Cecotti:2010fi}
\be\label{tttraces}
Z(q;n)=\mathrm{Tr}\Big(\mathbb{M}(q)^{n}\Big)
\ee
where $\mathbb{M}(q)$ is the 4d quantum monodromy \cite{Cecotti:2010fi,Cecotti:2014zga,Cecotti:2015qha} of the Argyres-Douglas SCFT of type $A_2$.
The traces in \eqref{tttraces} have been computed in \cite{Cecotti:2015lab} (see also \cite{Cecotti:2010fi}). They are not periodic in ${n}$ mod 5,
 although $Z(q;{n}+5)$ and $Z(q;{n})$ are expected to differ only by a simple factor.
 For ${n}>0$ the partition function has the asymptotic behavior \cite{Cecotti:2015lab}
 \be
 Z(e^{2\pi i\tau};{n})= \exp\!\left(\frac{2\pi i}{\tau} \frac{11}{60}\,{n}+O(1)\right)\quad
 \text{as }\tau\to 0,
 \ee
 which \emph{roughly speaking} says that the number of degrees of freedom living on the 
  longitudinal D3 probe is proportional to ${n}$. This shows that the dynamics of F-theory on a frozen 7-brane
 of type $(E_8,m=5)$ depends on $n$ in a very crucial way.
  
\medskip

We return to our transverse D3 probe wrapped on the last two factor spaces of \eqref{xxxxx65123}. 
In absence of better techniques to determine $h$ as a function of $n$, 
we revert to heuristic guessing. We stress again that $h$ is defined as the ``minimal'' number of everywhere light hypermultiplets which cannot be decoupled from the system (despite the fact that they look free in the IR) without introducing UV singularities.

 We still apply the image method and consider five transverse D3 branes
probing a 7-brane of type $\mathbf{II}$.
The 3-7 string states are not light for all values of $t$: their mass increases with the distance $|t|$ between the 7-brane and the D3 probes. In absence of the 7-brane the everywhere light 3-3 hypers
will arise from the $\cn=4$ $U(5)$ supermultiplet. Even in presence of the 7-brane
the center-of-mass of a system of transverse D3's
does not produce everywhere light states which cannot be eliminated
 since in the case of one D3 we get
the Argyres-Douglas model which has $h=0$.   We remain with the $SU(5)$ 3-3 degree of freedom.
In absence of 7-brane this would give 
\be
h_\text{no 7-brane}=\mathrm{rank}\,SU(5)=4.
\ee
However in presence of the 7-brane of type $\mathbf{II}$ two things may happen:
\begin{itemize}
\item[\it(i)] the interactions with the heavy 3-7 d.o.f.
may give mass to some hypermultiplet;
\item[\it(ii)] some light hypermultiplet may be non-minimal, i.e.\! can be decoupled without introducing UV singularities.
\end{itemize}

\textit{(i)} is unlikely to happen: at large distance from the 7-brane the local physics is the $\cn=4$ one,
and the hypermultiplets are BPS, hence their mass is protected against corrections due to
interactions with massive modes. On the contrary \textit{(ii)} refers to the regularity in the UV, in particular when the D3's approach the 7-brane. In this limit the difference between the different $n$'s is expected to be important by the argument around eq.\eqref{tttraces}. We also expect that $h$ is at least 1, since the partially frozen
geometries suffice to cover all known SCFT with $h=0$. Our heuristic guess leads to
\be\label{juyqwert}
1\leq h(n)\leq 4.
\ee
We have 4 possible values of $n\in (\Z/5\Z)^\times$, 4 possible values of $h(n)$, and no strong reason to believe that different values of $n$ lead to the same value of $h(n)$.
One may be tempted to think that there is a bijection between the 4 possible values of $n$
and the 4 values of $h(n)$ in the range \eqref{juyqwert}. Unfortunately there is no simple way to
decide the issue. As a matter of fact in the list of
known rank-1 SCFT in table 1 of \cite{M6} we have precisely one\footnote{\ We do not consider lightly-shadow entries because they are not \emph{bona fide} SCFTs, green one since have 12 supercharges,
and red ones since their existence is not confirmed.} model with $\Delta=6$
for each value \eqref{juyqwert} of $h(n)$ (besides the already discussed SCFTs with $h=0$ and $5$).
Thus \emph{potentially} we could have one distinct rank-1 special \emph{sub}-geometry\footnote{\ In the sense of
this paper.} per known rank-1 SCFT. 
The topic requires further analysis before making any definite assertion.

 \section*{Acknowledgments}
 I thank Mario Martone for sharing with me his insights about the physics
 at the discriminant locus, and Michele Del Zotto for correcting many
 of my misconceptions (the residual ones are my fault). I thank Yuji Tachikawa
 for clarifications about frozen singularities in M-theory.

\end{document}